\documentclass[twocolumn]{openjournal}
\usepackage{graphicx}
\usepackage{amsmath}
\usepackage{textcomp}
\usepackage{orcidlink}
\usepackage{hyperref}
\hypersetup{colorlinks=true, allcolors=[rgb]{0.0,0.0,0.5}
}

\shorttitle{Image-plane Lens Modeling of Interferometric Data}
\shortauthors{N. Zhang, et al.}

\begin{document}

\title{An Image-Plane Approach to Gravitational Lens Modeling of Interferometric Data}
\author{Nan Zhang$^{1}$\orcidlink{0000-0002-4861-0081}}\email{nanz6@illinois.edu}
\author{Sreevani Jarugula$^{2}$\orcidlink{0000-0002-5386-7076}}
\author{Justin~S.~Spilker$^{3}$\orcidlink{0000-0003-3256-5615}}
\author{Simon Birrer$^{4}$\orcidlink{0000-0003-3195-5507}}
\author{Jared Cathey$^{5}$\orcidlink{0000-0002-4657-7679}}
\author{Scott C.\ Chapman$^{6,7,8}$\orcidlink{0000-0002-8487-3153}}
\author{Veronica~J. Dike$^{9}$\orcidlink{0000-0002-9993-3796}}
\author{Anthony~H. Gonzalez$^{5}$\orcidlink{0000-0002-0933-8601}}
\author{Gilbert Holder$^{1}$\orcidlink{0000-0002-0463-6394}}
\author{Kedar~A. Phadke$^{9,10,11}$\orcidlink{0000-0001-7946-557X}}
\author{Cassie Reuter$^{12}$\orcidlink{0000-0001-7477-1586}}
\author{Joaquin~D. Vieira$^{9,10,1}$\orcidlink{0000-0001-7192-3871}}
\author{David Vizgan$^{9}$\orcidlink{0000-0001-7610-5544}}
\author{Dazhi Zhou$^{8}$\orcidlink{0000-0002-6922-469X}}

\affiliation{$^1$\text{Department of Physics, University of Illinois, 1110 West Green St., Urbana, IL 61801, USA}}
\affiliation{$^2$\text{Fermi National Accelerator Laboratory, Kirk Road and Pine Street, Batavia IL 60510-5011, USA}}
\affiliation{$^3$\text{Department of Physics and Astronomy and George P. and Cynthia Woods Mitchell Institute for Fundamental Physics}\linebreak\text{and Astronomy, Texas A\&M University, 4242 TAMU, College Station, TX 77843-4242, USA}}
\affiliation{$^4$\text{Department of Physics and Astronomy, Stony Brook University, Stony Brook, NY 11794, USA}}
\affiliation{$^5$\text{Department of Astronomy, University of Florida, Gainesville, FL 32611, USA}}
\affiliation{$^6$\text{Department of Physics and Atmospheric Science, Dalhousie University, Halifax, NS, B3H 4R2, Canada}}
\affiliation{$^7$\text{NRC Herzberg Astronomy and Astrophysics, 5071 West Saanich Rd, Victoria, BC, V9E 2E7, Canada}}
\affiliation{$^8$\text{Department of Physics and Astronomy, University of British Columbia, 6225 Agricultural Rd., Vancouver, V6T 1Z1, Canada}}
\affiliation{$^9$\text{Department of Astronomy, University of Illinois, 1002 West Green St., Urbana, IL 61801, USA}}
\affiliation{$^{10}$\text{Center for AstroPhysical Surveys, National Center for Supercomputing Applications, 1205 West Clark Street,}\linebreak \text{Urbana, IL 61801, USA}}
\affiliation{$^{11}$\text{NSF-Simons AI Institute for the Sky (SkAI), 172 E. Chestnut St., Chicago, IL 60611, USA}}
\affiliation{$^{12}$\text{Department of Physics, University of California, 366 Physics North MC 7300, Berkeley, CA, 94720-7300, USA}}

\begin{abstract}
Strong gravitational lensing acts as a cosmic telescope, enabling the study of the high-redshift universe. Astronomical interferometers, such as the Atacama Large Millimeter/submillimeter Array (ALMA), have provided high-resolution images of strongly lensed sources at millimeter and submillimeter wavelengths. To model the mass and light distributions of lensing and source galaxies from strongly lensed images, strong lens modeling for interferometric observations is conventionally performed in the visibility space, which is computationally expensive.
In this paper, we implement an image-plane lens modeling methodology for interferometric dirty images by accounting for noise correlations. We show that the image-plane likelihood function produces accurate model values when tested on simulated ALMA observations with an ensemble of noise realizations.
We also apply our technique to ALMA observations of two sources selected from the South Pole Telescope survey, comparing our results with previous visibility-based models. Our model results are consistent with previous models for both parametric and pixelated source-plane reconstructions. We implement this methodology for interferometric lens modeling in the open-source software package \texttt{lenstronomy}.
\end{abstract}

\section{Introduction}
Strong gravitational lensing is a powerful tool for studying cosmic history and the expansion of the universe. This phenomenon occurs when massive foreground galaxies or galaxy clusters distort the light from distant background objects. Lensing magnifies the flux of background galaxies, facilitating the study of high-redshift galaxies \citep{stark2008formation,jones2010resolved,treu2010strong,leethochawalit2016keck,spingola2020sharp}. Through observations and modeling of static and variable lensed objects, such as quasars, fundamental cosmological parameters, such as the Hubble constant, can be constrained \citep{birrer2020tdcosmo,millon2020tdcosmo}, and the distribution of dark matter can be mapped through gravitational imaging \citep{vegetti2010detection,vegetti2012gravitational,hezaveh2016detection,inoue2016alma,birrer2017lensing,minor2021unexpected,sengul2022substructure,csengul2022probing,ballard2024gravitational}. The flux-ratio statistics of strong lensing also currently provide the most stringent constraints on primordial black holes and the free-streaming properties of dark matter \citep{gilman2020warm,hsueh2020sharp,dike2023strong,keeley2024jwst}.

Submillimeter observations of high-redshift galaxies are important for studying the evolutionary history of early-type galaxies. Several gravitational lenses have been discovered at millimeter or submillimeter wavelengths with the \textit{Herschel}-ATLAS survey \citep{eales2010herschel}, Atacama Cosmology Telescope \citep{marsden2014atacama}, Planck \citep{planck2013results}, and the South Pole Telescope (SPT) \citep{vieira2013dusty} and have been followed up with Atacama Large Millimeter/submillimeter Array (ALMA) observations, which provide higher sensitivity and resolution \citep{hezaveh2013alma,vlahakis20152014,spilker2016alma,berman2022passages}.

Lens modeling techniques are needed to extract information from observed images of strongly lensed galaxies. The physical properties of both the lens and source galaxies are inferred statistically through lens modeling. This is achieved by assuming that the lensed images can be traced back to the flux from the source galaxies, while the light-path deflection angles are determined by the density distribution of the lens galaxy.

To estimate lens models that best fit strongly lensed images and provide reasonable statistical descriptions, good noise models of data images are needed.
However, while interferometers such as ALMA provide high resolution, the noise in interferometric images is highly correlated because signals are sampled in Fourier space, known as the $uv$-plane. The image-space noise correlation makes it more challenging to analyze strong lensing systems observed by interferometers. Several lens modeling packages have been developed to address this challenge. Examples include \cite{spilker2016alma}, \cite{hezaveh2016detection}, \cite{powell2021novel}, and \cite{nightingale2021pyautolens}, which focus on lens modeling in visibility space. Recently, \cite{powell2022lensed} proposed a ``fast $\chi^2$'' technique that enables the computation of $\chi^2$ using only dirty images and a sub-grid of the $uv$-plane, providing an image-plane approach to lens modeling of interferometric images.

In this paper, we re-derive the mathematical description of the fast $\chi^2$ method and develop an independent implementation within the widely-used lenstronomy analysis package. The resulting likelihood function enables image-plane lens modeling of interferometric images. The lens modeling computation requires only three images of an interferometric observation as input: naturally weighted data images (also known as dirty images), the corresponding dirty beam response maps, which we refer to as the Point Spread Function (PSF), and the primary beam image of the data. We validate this image-plane lens modeling approach using simulated ALMA images, ensuring that the image-plane likelihood function yields the expected statistical properties. We also perform fittings on real ALMA images to verify that the results of image-plane-based fittings align with previous $uv$-fitting results obtained from the same data.

We integrate our technique into the open-source \texttt{Lenstronomy} package \citep{birrer2018lenstronomy,Birrer2021}. \texttt{Lenstronomy} includes a large library of parametric lens and source profiles for parametric fitting. It also provides auxiliary tools for computing mappings between source planes and image planes, enabling pixelated source reconstructions. With the integrated capability to fit interferometric images, the multiband fitting feature of \texttt{Lenstronomy} also facilitates the joint analysis of strong lensing images from both interferometers like ALMA and pixel-based imaging telescopes, including the Hubble Space Telescope and the James Webb Space Telescope.

This paper is organized as follows. In Section \ref{sec methodology}, we introduce the image-plane natural-weighting likelihood function and present the method for pixelated source reconstruction using the natural-weighting covariance matrix. In Section \ref{sec test likelihood}, we evaluate the statistical properties of the image-plane likelihood function by fitting a set of simulated ALMA data with an ensemble of random noise realizations. We also compare image-plane fitting with $uv$-plane fitting to verify the validity of our method. In Section \ref{sec fitting alma data}, we test our method with two well-studied ALMA sources. Parametric fitting is performed on SPT0346-52 and compared with previous analyses based on the $uv$-plane fitting package \texttt{visilens} \citep{hezaveh2013alma,spilker2016alma,dong2019source}. We test pixelated source reconstruction on SPT0311-58, which was previously analyzed using the $uv$-plane pixelated source fitting code \texttt{RIPPLES} \citep{hezaveh2016detection,spilker2022chaotic}. Finally, we summarize our work in Section \ref{Sec discussion}.

\section{Methodology of Lens modeling}
\label{sec methodology}

\begin{figure*}[htp]
\centering
\includegraphics[trim={0.0cm 0.0cm 0.0cm 0.0},clip,width=0.95\textwidth]{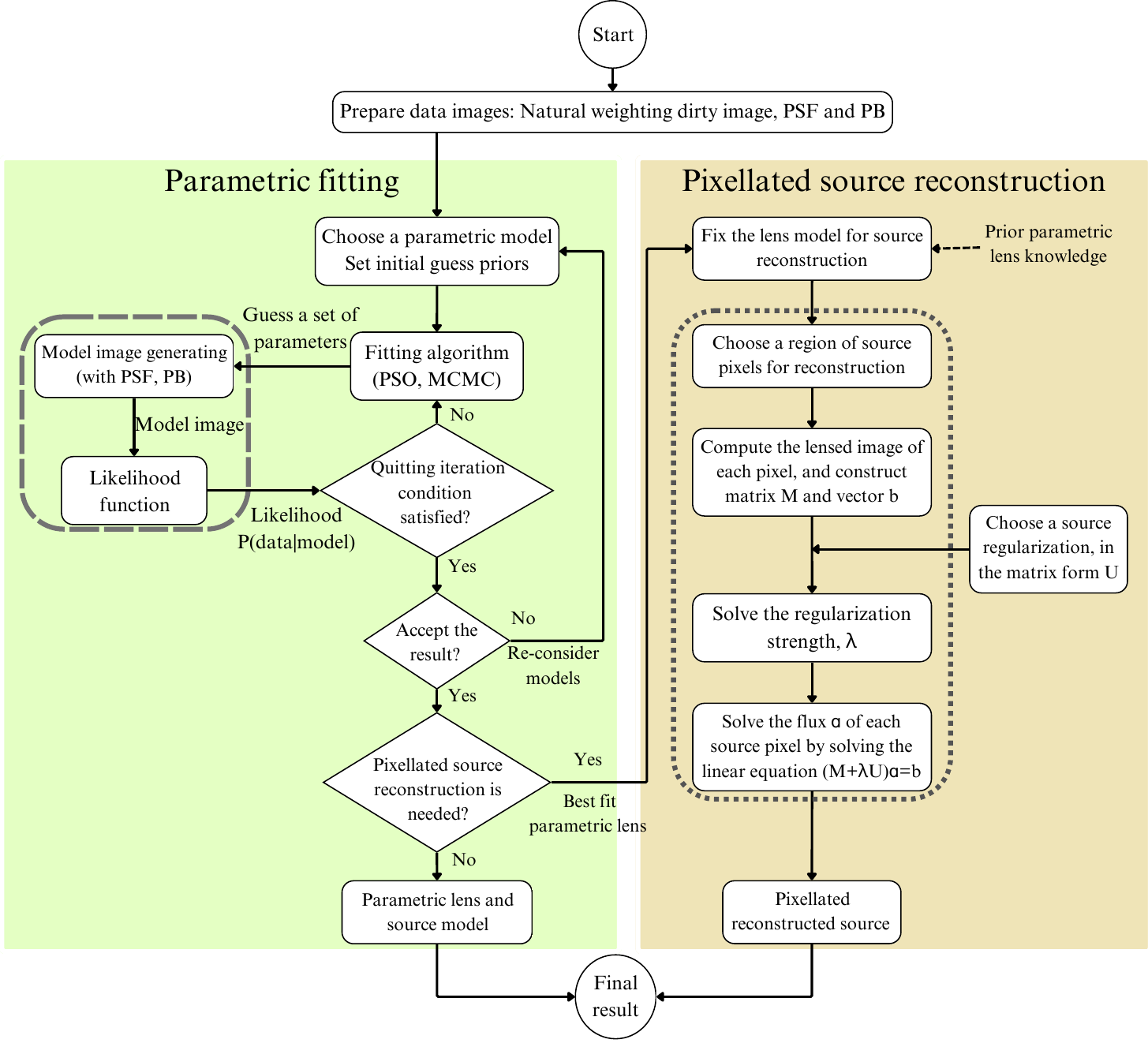}
\caption{Flowchart of the lens modeling pipeline for interferometric images. The pipeline consists of two components: natural-weighting parametric fitting (left-hand panel) and pixelated source reconstruction (right-hand panel). Our method takes as input a set of three data files from interferometric imaging: the naturally weighted dirty image, the dirty beam (Point Spread Function, PSF), and the primary beam (PB). The likelihood function (Equation \ref{likelihood 1}) is maximized using optimization algorithms such as Particle Swarm Optimization (PSO) and Markov Chain Monte Carlo (MCMC). The long-dashed box (left-hand side) highlights modifications made to the fitting pipeline of \texttt{Lenstronomy} to accommodate interferometric fitting. Pixelated source reconstruction is performed with a fixed parametric lens model, which can either be the best-fit parametric lens model from previous analysis or a model from the literature, referred to as ``prior parametric lens knowledge.'' The source reconstruction pipeline is detailed within the dashed box on the right-hand side. The mathematical symbols used in this pipeline are introduced and explained in Section \ref{sec source recon equations}. Our model outputs a parametric lens model with a parametric source (from parametric fitting) or a pixelated reconstructed source (from source reconstruction). The pixelated source reconstruction integrated into \texttt{Lenstronomy} can also be applied to non-interferometric data, such as HST or JWST images, if the likelihood function used for interferometric data is replaced with the appropriate likelihood function for non-interferometric data.
}
\label{fig flow chart}
\end{figure*}

In this section, we present our lens modeling method for interferometric images, which consists of two components: parametric fitting and linear pixelated source reconstruction. To illustrate how our methods are integrated into the full analysis of lens modeling, we summarize our pipeline in Figure \ref{fig flow chart}. In Section \ref{Sec interferometry imaging}, we provide an overview of interferometric visibility measurements and image-plane maps of an observation. In Section \ref{sec stat functions fitting}, we introduce the natural-weighting likelihood function for image-plane lens modeling. In Section \ref{sec source recon equations}, we present the method for pixelated source reconstruction, which utilizes the natural-weighting covariance matrix.

\subsection{Interferometry Imaging and Noise}
\label{Sec interferometry imaging}
In this section, we briefly introduce interferometric visibilities and images with the associated noise properties. For a more general introduction to interferometry, we refer readers to \cite{thompson2017interferometry}.

\subsubsection{The Sky Image with the Primary Beam}
We denote the true sky image as ${x}_{\rm{true}}(\boldsymbol{\theta})$, which describes the distribution of the light intensity at the wavelength of the observation. We discretize the sky image into a pixelated image vector $\boldsymbol{x}_{\rm{true},i} = {x}_{\rm{true}}(\boldsymbol{\theta_i})$, where $\boldsymbol{\theta_i}$ denotes the two-dimensional coordinate of the sky at the $i$-th image pixel.

One complication in observing the true sky with interferometric antennas is the antenna's \emph{primary beam}. The primary beam characterizes the angular sensitivity of the antenna's observation of the sky. This sensitivity arises from the diffraction of signal radio waves due to the shape of the antenna. Hence, the signal sky image observed by the telescope is not the true sky image, but the product of the true sky image and the antenna primary beam:
\begin{equation}
\label{primary beam multiplication0}
{x}(\boldsymbol{\theta}) = A(\boldsymbol{\theta}) {x}_{\rm{true}}(\boldsymbol{\theta}).
\end{equation}
We use the discretized image vector $\boldsymbol{x}_i=x(\boldsymbol{\theta_i})$ to express the signal image containing the effect of the primary beam. The effect of the primary beam is then encoded in the matrix $\mathbf{A}$:
\begin{equation}
\label{primary beam multiplication}
\boldsymbol{x} = \mathbf{A} \boldsymbol{x}_{\rm{true}},
\end{equation}
where $\mathbf{A}$ acts on the true sky image vector. The primary beam itself is an image of the same size as the true sky image. The matrix $\mathbf{A}$ acts as the pixel-wise multiplication of the true sky image and the primary beam image. Therefore, $\mathbf{A}$ is a diagonal matrix, with diagonal entries $\mathbf{A}_{ii} = A(\boldsymbol{\theta}_i)$, where $A(\boldsymbol{\theta}_i)$ is the value of the primary beam map at the image space coordinate $\boldsymbol{\theta}_i$.

The primary beam will be applied whenever a sky model image is generated. Therefore, the dirty image input for the fitting code should not be previously primary-beam corrected by imaging software like the \emph{Common Astronomy Software Applications} (\texttt{CASA}) \citep{mcmullin2007casa,bean2022casa}.

\subsubsection{Visibility Measurements and the Visibility Noise}
An interferometer is an array of antennas that measures the sky brightness in Fourier space, conventionally referred to as the $uv$-plane or visibility space. According to the van Cittert-Zernike theorem, the correlation of signals from a pair of antennas (a single \emph{baseline}) samples a pair of visibilities in the $uv$-plane \citep{thompson2017interferometry}.
Not considering noise, the signal visibility is the Fourier transform of the signal image:
\begin{equation}
\label{signal vis from signal image}
v_s(\boldsymbol{u}) = \int_{-\infty}^{\infty}x(\boldsymbol{\theta})e^{-2\pi i \boldsymbol{u}\cdot\boldsymbol{\theta}}d\boldsymbol{\theta}.
\end{equation}
where the signal sky image is given by Equation (\ref{primary beam multiplication0}).

We denote the measured visibility as $v(\boldsymbol{u}_l)$, where $\boldsymbol{u}_l$ denotes the coordinate of the $l$-th measured visibility in the $uv$-plane. The value of $\boldsymbol{u}_l$ is the physical distance between the two antennas of a baseline, measured in wavelengths. The measured visibility contains the signal visibility and noise:
\begin{equation}
\label{visibility expand}
v(\boldsymbol{u}_l) = v_s(\boldsymbol{u}_l) + v_{n}(\boldsymbol{u}_l),
\end{equation}
where $v_{n}(\boldsymbol{u}_l)$ is the random noise of the visibility measurement.

In real observations, there is another aspect of the visibility noise known as the phase and amplitude errors. The phase and amplitude errors are introduced along the path of the radio waves from the source to the observer. Since they are correlated with the signal visibility, most phase and amplitude errors can be solved in the data-processing procedure through self-calibration, thereby reducing sensitivity loss \citep{thompson2017interferometry}. We choose not to address antenna-based phase and amplitude errors in this work. In this analysis, we refer to the term `noise' only as the random noise of the complex visibility (i.e., the second term of Equation \ref{visibility expand}). \cite{hezaveh2013alma} combined phase error fitting with lens modeling in the $uv$-plane pipeline, calibrating the phase errors and computing the likelihood of lens modeling iteratively.

\subsubsection{Data Image, the Image-plane Noise, and the Dirty Beam}
To make the data image, the measured visibility is inversely Fourier transformed to the image plane. \emph{Visibility weights} are introduced to be multiplied onto the visibility during the transformation process. Denoting $w_l$ as the weighting factor for the $l$-th visibility, the data image $d$ is computed as
\begin{equation}
\label{def data image}
d(\boldsymbol{\theta}) = \frac{1}{W}\sum_l  w_l\Big[v(\boldsymbol{u}_l)e^{2\pi i\boldsymbol{u}_l\cdot\boldsymbol{\theta}}+v^*(\boldsymbol{u}_l)e^{-2\pi i\boldsymbol{u}_l\cdot\boldsymbol{\theta}}\Big].
\end{equation}
The summation $\sum_l$ is over the measured visibilities. $W$ in Equation (\ref{def data image}) is a normalization factor determined by the normalization scheme used when making the images. We choose to use the ``Sum-Of-Weights''\footnote{\url{https://casadocs.readthedocs.io/en/stable/notebooks/synthesis_imaging.html\#Normalization}} normalization. In our notation, it corresponds to
\begin{equation}
\label{W def}
W=2\sum_l w_l.
\end{equation}

The PSF in the image plane is the inversely Fourier-transformed weights. Therefore,
\begin{equation}
\label{def A psf matrix}
B(\boldsymbol{\theta}_i,\boldsymbol{\theta}_j) = \frac{1}{W}\sum_{l}  w_l \Big[ e^{2\pi i \boldsymbol{u}_l\cdot(\boldsymbol{\theta}_i - \boldsymbol{\theta}_j)} + e^{-2\pi i \boldsymbol{u}_l\cdot(\boldsymbol{\theta}_i - \boldsymbol{\theta}_j)}\Big],
\end{equation}
where the two positional coordinates $\boldsymbol{\theta}_i$ and $\boldsymbol{\theta}_j$ indicate that the matrix entry $\mathbf{B}_{ij} = B(\boldsymbol{\theta}_i,\boldsymbol{\theta}_j)$ represents the contribution of the flux from the $i$-th pixel of the unconvolved image to the $j$-th pixel of the convolved image. This PSF is also known as the \emph{dirty beam} response in the interferometric literature.

To conclude, with the effects of the primary beam (Equation \ref{primary beam multiplication}), the PSF (Equation \ref{def A psf matrix}), and the noise, the interferometric data image (Equation \ref{def data image}) is now written using image-plane maps exclusively as
\begin{equation}
\boldsymbol{d} = \mathbf{B}\mathbf{A}\boldsymbol{x}_{\rm{true}} + \boldsymbol{n}.
\end{equation}

\subsubsection{Noise Covariance Matrix in the Image Plane under Natural Weighting}
To infer physical parameters from the data image, it is also necessary to know the noise covariance matrix in the image plane. To derive this image-plane noise covariance matrix, we start with the product of noise values at two pixels. The product of the noise at two pixels is given by
\begin{equation}
\label{noise cov matrix fourier derivation0}
\begin{aligned}
n(\boldsymbol{\theta}_i)n(\boldsymbol{\theta}_j)=&\frac{1}{W^2}\sum_{l_1}\sum_{l_2}w_{l_1}w_{l_2}\\
\cdot&\Big[v_{{n}}(\boldsymbol{u}_{l_1})e^{2\pi i \boldsymbol{u}_{l_1}\cdot\boldsymbol{\theta}_i} + v^*_{{n}}(\boldsymbol{u}_{l_1})e^{-2\pi i \boldsymbol{u}_{l_1}\cdot\boldsymbol{\theta}_i}\Big]\\
\cdot&\Big[v^*_{{n}}(\boldsymbol{u}_{l_2})e^{-2\pi i \boldsymbol{u}_{l_2}\cdot\boldsymbol{\theta}_j} + v_{{n}}(\boldsymbol{u}_{l_2})e^{2\pi i \boldsymbol{u}_{l_2}\cdot\boldsymbol{\theta}_j}
\Big].
\end{aligned}
\end{equation}
Note that we have expanded the noise of the $j$-th pixel, $n(\boldsymbol{\theta}_j)$, into its complex conjugate. This does not alter the equation since the image pixel values are real. We then take the ensemble average of the noise product, denoted by $\langle\dots\rangle$. The image-plane noise covariance is
\begin{equation}
\label{noise cov matrix fourier derivation}
\begin{aligned}
\langle n(\boldsymbol{\theta}_i)n(\boldsymbol{\theta}_j)\rangle
= &\frac{1}{W^2}\sum_{l} w_{l}^2 \langle|v_{{n}}(\boldsymbol{u}_l)|^2\rangle
\\&\cdot\Big[ e^{2\pi i \boldsymbol{u}_l \cdot(\boldsymbol{\theta}_i - \boldsymbol{\theta}_j)} + e^{-2\pi i \boldsymbol{u}_l \cdot(\boldsymbol{\theta}_i - \boldsymbol{\theta}_j)}
\Big],
\end{aligned}
\end{equation}
where the cross terms of different visibilities, $\langle v_{{n}}(\boldsymbol{u}_{l_1})v^*_{{n}}(\boldsymbol{u}_{l_2})\rangle$ (for $l_1\neq l_2$), vanish because the random noise in each visibility is independent.

To make the computation of the likelihood function feasible, we choose the visibility weights ($w_l$) to be in the \emph{natural weighting}, where the visibility weights are the inverse of the visibility variance, i.e.,
\begin{equation}
\label{natwt}
w_l = \frac{1}{\langle|v_{{n}}(\boldsymbol{u}_l)|^2\rangle}.
\end{equation}
With natural weighting, the Fourier coefficients of the noise covariance (Equation \ref{noise cov matrix fourier derivation}) simplify to just $w_l$, as the visibility variance cancels out with one weighting factor. The entries of the noise covariance matrix are then computed as
\begin{equation}
\begin{split}
\label{noise cov matrix fourier derivation2}
(\mathbf{C}_n)_{ij}&=\langle n(\boldsymbol{\theta}_i)n(\boldsymbol{\theta}_j)\rangle \\ &= \frac{1}{W^2}\sum_{l} w_l
\Big[ e^{2\pi i \boldsymbol{u}_l \cdot(\boldsymbol{\theta}_i - \boldsymbol{\theta}_j)} + e^{-2\pi i \boldsymbol{u}_l \cdot(\boldsymbol{\theta}_i - \boldsymbol{\theta}_j)}
\Big].
\end{split}
\end{equation}
The right-hand side of the covariance matrix above is proportional to the natural weighting PSF (Equation \ref{def A psf matrix}) by a factor of $1/W$.

Under our normalization scheme given by Equation (\ref{W def}), the diagonal entries of the matrix $\mathbf{B}$ are normalized to one. On the other hand, the diagonal entries of $\mathbf{C}_n$ are the noise autocorrelation of each pixel. We denote these diagonal entries of the noise covariance matrix as $\sigma_n^2$. Hence, the proportionality between the noise covariance matrix and the PSF can also be expressed as
\begin{equation}
\label{noise covariance psf proportionality}
\mathbf{B} = \frac{1}{\sigma_n^2}\mathbf{C}_n.
\end{equation}

The simple relationship between the noise covariance matrix and the PSF matrix holds only for naturally weighted dirty images. This relationship does \emph{not} apply in two important cases. First, for other weighting schemes, such as robust weighting, which may enhance small-scale details in the image, the proportionality relation indicated by Equation (\ref{noise covariance psf proportionality}) no longer holds. Additionally, this relationship does not apply to ``cleaned'' images, even though such images, generated by the CLEAN algorithm, are typically preferred in interferometric image analyses.

\subsubsection{uv Gridding and FFT}
We have assumed that the transformation between the $uv$-plane and the image plane is a direct Fourier transform (DFT) in the above calculations. However, in practice, the densely and irregularly distributed visibility is regridded into a regularly spaced grid in order to perform a fast Fourier transform (FFT). For example, \texttt{CASA} regrids the weighting factors and weighted visibilities before generating the images. During the $uv$ gridding, as the dense visibility is reduced and mapped onto a regular grid, the information in the data is compressed, which is non-reversible. Hence, $uv$ gridding has the potential to cause errors in the analysis, especially for the image-plane approach used in this work. However, our simulation test in Section \ref{sec test likelihood} shows no obvious negative effects caused by $uv$ gridding in the lens modeling.

\subsection{Statistical Inference of Lens and Source Models.}
\label{sec stat functions fitting}
Strong lens modeling aims to find the best-fit models for lens and source galaxies that reproduce the observed lensed images. Bayesian statistics quantitatively determine how well candidate models fit the data. In this section, we apply image-plane lens modeling within a Bayesian framework, where the likelihood function fully accounts for the noise covariance. We further extend this approach in Section \ref{sec source recon equations}, introducing the method of pixelated source reconstruction with source priors, combined with the image-plane likelihood function.

\subsubsection{Images of Strong Lensing}
Gravitational lensing can be regarded as a mapping of surface brightness from the source plane to the image plane. The physical parameters of interest can be cast into two categories: lens parameters and source parameters. Denoting the parameter set as $p$, we write $p = \{p_{lens},p_{source}\}$. We express the true sky flux image introduced in Equation (\ref{primary beam multiplication}) as a function of these parameters:
\begin{equation}
\label{lens linear map x0=La}
\boldsymbol{x}_{\rm{true}}(p) = \hat{\mathbf{L}}(p_{lens})\boldsymbol{s}(p_{source}),
\end{equation}
where $\hat{\mathbf{L}}$ is the lensing operator acting on the source surface brightness, denoted by $\boldsymbol{s}$. Admitting the primary beam effect in interferometric images, we define the unconvolved image $\boldsymbol{x}(p)$ as the ``model'' image:
\begin{equation}
\label{lens linear map x=La}
\boldsymbol{x}(p) = \mathbf{A}\hat{\mathbf{L}}(p_{lens})\boldsymbol{s}(p_{source}).
\end{equation}

\subsubsection{Natural Weighting Likelihood Function}
\label{sec natwt likelihood}
Bayes' Theorem states that the probability distribution of a model given the data (posterior) is the normalized product of the probability of the data given the model (likelihood) and the prior knowledge of the model:
\begin{equation}
\label{bayes formula}
P\big[\boldsymbol{x}(p)\big|\boldsymbol{d}\big] = \frac{P\big[\boldsymbol{d}\big|\boldsymbol{x}(p)\big]P\big[\boldsymbol{x}(p)\big]}{P(\boldsymbol{d})}.
\end{equation}
The general form of the likelihood function is
\begin{equation}
\label{likelihood 1}
P(\boldsymbol{d}|\boldsymbol{x}) = \frac{1}{Z_d}\exp\Big[{-\frac{1}{2} (\boldsymbol{d} - \mathbf{B}\boldsymbol{x})^T \mathbf{C}_{n}^{-1}(\boldsymbol{d} - \mathbf{B}\boldsymbol{x})}\Big].
\end{equation}
For interferometric images, the matrix $\mathbf{B}$ represents the dirty beam response, which acts as the PSF convolution on the model image. The product $\mathbf{B}\boldsymbol{x}$ is the dirty image, and $(\boldsymbol{d} - \mathbf{B}\boldsymbol{x})$ gives the \emph{residuals} of the dirty image. The matrix $\mathbf{C}_n$ is the noise covariance matrix in the image plane. Since the normalization factor $Z_d = \sqrt{(2\pi)^{N_d}\det{\mathbf{C}_n}}$ depends only on the noise covariance matrix and the data dimension $N_d$, maximizing the likelihood function is equivalent to minimizing the $\chi^2$ function, defined as
\begin{equation}
\label{chi2 def from likelihood}
\chi^2 = (\boldsymbol{d} - \mathbf{B}\boldsymbol{x})^T \mathbf{C}_{n}^{-1}(\boldsymbol{d} - \mathbf{B}\boldsymbol{x}).
\end{equation}

In lens modeling of interferometric images, the challenging part of computing the likelihood function is correctly accounting for the noise covariance matrix $\mathbf{C}_n$. Under the natural weighting condition, incorporating the noise covariance into the likelihood function is significantly simplified. Substituting Equation (\ref{noise covariance psf proportionality}) into Equation (\ref{chi2 def from likelihood}) and expanding the brackets, $\chi^2$ is computed as
\begin{equation}
\begin{aligned}
\label{chi2 simplified}
\chi^2  &= \frac{1}{\sigma_n^2}(\boldsymbol{d}-\mathbf{B}\boldsymbol{x})^T\mathbf{B}^{-1}(\boldsymbol{d}-\mathbf{B}\boldsymbol{x})\\
&= \frac{1}{\sigma_n^2} (\boldsymbol{d}^T\mathbf{B}^{-1}\boldsymbol{d} - 2\boldsymbol{d}^T\boldsymbol{x} + \boldsymbol{x}^T \mathbf{B}\boldsymbol{x}).
\end{aligned}
\end{equation}
The second line of Equation (\ref{chi2 simplified}) is the $\chi^2$ function used our lens modeling code. The model image vector $\boldsymbol{x}$ in the $\chi^2$ computation has already accounted for the effect of the primary beam, as shown in Equation (\ref{lens linear map x=La}).

This simplified $\chi^2$ formulation makes lens modeling computationally feasible, benefiting from the cancellation of the inverse covariance matrix and the PSF convolution matrix. The most computationally expensive step in the $\chi^2$ calculation (Equation \ref{chi2 simplified}) is the convolution of the PSF with the model, i.e., $\mathbf{B}\boldsymbol{x}$, which can be efficiently computed using an FFT-based convolution algorithm. This step has a time complexity of $O(N_d\ln(N_d))$. In contrast, for the general $\chi^2$ defined in Equation (\ref{chi2 def from likelihood}), matrix-vector multiplications, such as the multiplication of the noise covariance matrix $\mathbf{C}_n^{-1}$ with the residual vector $(\boldsymbol{d}-\mathbf{B}\boldsymbol{x})$, require $O(N_d^2)$ operations. Thus, the simplified-$\chi^2$ method significantly reduces computational time compared to the general $\chi^2$ function.

Although the first term of the simplified $\chi^2$, $\boldsymbol{d}^T\mathbf{B}^{-1}\boldsymbol{d}$, requires inverting the convolution matrix and performing matrix-vector multiplications, this term remains constant throughout the fitting process and does not affect the fitting results. Therefore, it does not need to be computed at all.

\subsubsection{Parametric Fitting with Flat Priors}
\label{sec para fitting flat prior}
Parametric fitting involves performing lens modeling using pre-defined profile functions that describe the mass density of lens galaxies and the flux of both lens and source galaxies. With the image-plane $\chi^2$ (Equation \ref{chi2 simplified}), it is straightforward to use optimization algorithms such as Particle Swarm Optimization (PSO) to search for best-fit parametric models by maximizing the likelihood function over parameter spaces. It is also possible to sample the posterior distribution of the parameters using MCMC.

We choose to use flat priors for parameters within finite search limits. In Section \ref{sec test likelihood}, we will examine the statistical behavior of the image-plane $\chi^2$ (Equation \ref{chi2 simplified}) in parametric fitting, while also exploring the biases introduced by the flat sampling priors.

\subsection{Source Plane Reconstruction with Regularizing Source Priors}
\label{sec source recon equations}
In this section, we focus on a specific problem of lens modeling inference, where the light flux distribution of background source galaxies is described by a linear combination of source light image vectors.

While parametric fitting is a powerful tool for identifying strong lensing models with pre-defined profiles, these profiles cannot fully capture the complexity of real astrophysical sources. Specifically, the long baselines of radio interferometers enable sub-arcsecond imaging, revealing fine galactic structures that simple source profiles cannot adequately describe. In such cases, parametric fitting alone is insufficient for reconstructing the detailed morphology of source galaxies.

One method to increase the flexibility of source profiles is to use a more complete set of source light functions for modeling source brightness, such as pixelated sources with uniform or adaptive grids \citep{vegetti2009bayesian,nightingale2015adaptive}. This approach is known as ``semi-linear'' \citep{warren2003semilinear} because the source parameters, which are the amplitudes of each source profile, are inferred linearly while the non-linear lens parameters are fixed. Thus, a key step in semi-linear inversion is solving for the source amplitudes given a fixed lens model.

To avoid overfitting in the semi-linear source reconstruction, \emph{source regularization} is also necessary. It ensures that the reconstructed source morphology remains smooth by preventing abrupt flux variations between neighboring source pixels. Regularization enters the Bayesian framework of strong lens modeling as a prior probability distribution of the source parameters \citep{suyu2006bayesian}.

The process of pixelated source reconstruction for interferometric strong lensing images follows the general principles of semi-linear inversion, with the key distinction that the likelihood function used is the natural weighting likelihood function introduced in Section \ref{sec natwt likelihood}. Since this likelihood function is based on natural weighting, the source reconstruction must also be performed on naturally weighted images. Figure \ref{fig flow chart} illustrates the pipeline for pixelated source reconstruction. Alongside the flowchart, we describe the motivation and mathematical framework, using interferometric notation, for each step of the reconstruction process.

\subsubsection{Linear Expansions of Source and Image Vectors}
First, we express the source-plane image as a linear combination of a set of light profile vectors:
\begin{equation}
\label{s = a_i s_i}
\boldsymbol{s} = \sum_{i} a_i \boldsymbol{s}_i.
\end{equation}
Here, $\boldsymbol{s}$ on the left-hand side is the discretized source image vector, while each $\boldsymbol{s}_i$ on the right-hand side corresponds to a source profile in which only a single pixel (the $i$-th pixel) carries a unit flux. With this interpretation, we refer to the vectors $\boldsymbol{s}_i$ simply as source pixels.

For pixelated source reconstruction, the source pixels are chosen to cover the background source region of interest. The linear coefficients $\{a_i\}$ physically represent the flux of each source pixel.
Applying the source expansion (Equation \ref{s = a_i s_i}), the lensed model image (Equation \ref{lens linear map x=La}) can then be written as
\begin{equation}
\label{x = AL a_i s_i}
\boldsymbol{x} =\mathbf{A}\hat{\mathbf{L}}(\sum_{i}a_i \boldsymbol{s}_i).
\end{equation}
When the model of the lensing galaxy is fixed, the amplitudes of the source pixels remain linear with respect to the lensed images, allowing the linear coefficients to be factored out of the lensing operator:
\begin{equation}
\label{x = a_i AL s_i}
\boldsymbol{x} =\sum_{i}a_i \mathbf{A}\hat{\mathbf{L}}\boldsymbol{s}_i.
\end{equation}
The collection of the lensed source pixels forms a set of image-plane vectors, which we refer to as the ``lensed vector set''.
The $i$-th lensed vector is denoted as
\begin{equation}
\label{x_i def}
\boldsymbol{x}_i = \mathbf{A}\hat{\mathbf{L}}\boldsymbol{s}_i,
\end{equation}
which corresponds to the lensed image of the $i$-th source pixel.
Thus, the lensed image becomes a linear combination of lensed vectors:
\begin{equation}
\label{x = a_i x_i}
\boldsymbol{x}=\sum_{i}a_i \boldsymbol{x}_i.
\end{equation}

\subsubsection{Image-plane $\chi^2$ as a Quadratic Function of Source Amplitudes}
With the lensed image expressed as a linear combination, we compute the $\chi^2$ function in terms of the linear coefficients $\{a_i\}$ by applying the lensed image expansion (Equation \ref{x = a_i x_i}) to the image-plane $\chi^2$ (Equation \ref{chi2 simplified}). The result is
\begin{equation}
\label{source plane recon chi2 expanded}
\chi^2 = \frac{1}{\sigma^2_n}(\boldsymbol{d}^T\mathbf{B}^{-1}\boldsymbol{d})- 2 \sum_{i}a_i b_i+\sum_{i,j}a_ia_j M_{ij},
\end{equation}
where
\begin{equation}
\label{source recon b}
b_i = \frac{\boldsymbol{d}^T\boldsymbol{x}_i}{\sigma^2_n},
\end{equation}
and
\begin{equation}
\label{source recon m}
M_{ij} = \frac{\boldsymbol{x}_i^T\mathbf{B}\boldsymbol{x}_j}{\sigma^2_n}.
\end{equation}
Since the $\chi^2$ here is a quadratic function of the source amplitudes $\{a_i\}$, the best-fit amplitudes can be obtained, in the absence of a regularization term, by solving a linear system using standard linear algebra techniques.

\subsubsection{The Source Regularization Term}
For source reconstruction without regularization, there is a risk of overfitting, where the amplitudes of neighboring source pixels oscillate drastically to fit the lensed image.
To prevent overfitting, a penalty term must be introduced to suppress excessive oscillations among neighboring source pixels while optimizing $\chi^2$. This penalty term is known as source regularization. In the Bayesian framework, the regularization term serves as the prior function for the source parameters. Commonly used regularization methods for pixelated source reconstruction include zeroth-order, gradient, and curvature regularization, as introduced by \cite{suyu2006bayesian}. These regularization schemes mitigate overfitting by either suppressing pixels with large flux values (zeroth-order regularization) or penalizing large flux differences between neighboring pixels (gradient and curvature regularization).

The regularization schemes mentioned above are incorporated into the Bayesian prior function in the general form:
\begin{equation}
\label{regularization general form}
P(\{a_i\}|\lambda_s,U) =\frac{1}{Z_{a}(\lambda_s, U)} \exp(-\frac{1}{2}\lambda_s \sum_{ij}U_{ij}a_ia_j).
\end{equation}
In this expression, the number $\lambda_s$ and the matrix $U$ parameterize the source regularization. The form of $U_{ij}$ determines the type of regularization, while the overall strength of the regularization is controlled by $\lambda_s$. Similar to $\chi^2$, the exponent of the source regularization prior, as written above, is a quadratic function of the source amplitudes $\{a_i\}$, facilitating the source inference described below.

The computation of the zeroth-order, gradient, and curvature regularization matrices for pixelated sources is also provided by \cite{suyu2006bayesian}. In Appendix \ref{sec appendix regularization}, we present our method for computing these regularization matrices, leveraging the fact that they exhibit block-matrix structures.

\subsubsection{Source Amplitude Inference with the Regularization Term}

Given the image-plane $\chi^2$ (proportional to the logarithm of the likelihood, Equation \ref{likelihood 1} and Equation \ref{source plane recon chi2 expanded}) and the source regularization prior (Equation \ref{regularization general form}), the logarithm of the posterior is written as
\begin{equation}
\label{posterior source recon}
\begin{split}
\ln\Big[P(\{a_i\},\boldsymbol{d}|\lambda_s, U)\Big] = &-\frac{1}{2}\Big[\sum_{ij}(\lambda_s U_{ij}+M_{ij})a_ia_j \\
&- 2\sum_{i}a_ib_i
+ \frac{1}{\sigma^2_n}\boldsymbol{d}^T\mathbf{B}^{-1}\boldsymbol{d}\Big]\\
&- \ln{Z_d} - \ln{Z_a(\lambda_s, U)}.
\end{split}
\end{equation}
The last line of the log-posterior contains the normalization terms for the likelihood and prior functions, respectively, which do not change with respect to the source amplitudes $\{a_i\}$. The optimal $\{a_i\}$, which maximizes the logarithm of the posterior, is solved linearly from
\begin{equation}
\label{linear amp solving equation with prior}
\sum_{j}(\lambda_s U_{ij}+M_{ij})a_j = b_i.
\end{equation}
Substituting the solution for $\{a_i\}$ back into the source-plane expansion (Equation \ref{s = a_i s_i}) and image expansion (Equation \ref{x = a_i x_i}) yields the most probable source and lensed images.

\subsubsection{Amplitude-marginalized Posterior for Ranking Lens Models and Specifying Regularization Strength}
For the source amplitude inference introduced above, we need to specify the regularization strength $\lambda_s$. One option is to use the $\hat{\lambda}_s$ that maximizes the amplitude-marginalized posterior function. Additionally, the amplitude-marginalized posterior can also be used for ranking lens models \citep{suyu2006bayesian}.

Because the image-plane $\chi^2$ and the exponent of the source regularization prior are quadratic functions of source amplitudes, we can analytically integrate the posterior function over the amplitude parameter subspace. The result is the amplitude-marginalized posterior:
\begin{equation}
\label{source recon marginalized post}
\begin{split}
P(\boldsymbol{d}|\lambda_s,U) = &\frac{\exp(-\frac{\boldsymbol{d}^T\mathbf{B}^{-1}\boldsymbol{d}}{2\sigma^2_n})}{Z_d}\sqrt{\frac{\det(\lambda_s U)}{\det(M+\lambda_s U)}}
\\&
\cdot \exp\Big[\frac{1}{2}b^T(M+\lambda_s U)^{-1}b\Big].
\end{split}
\end{equation}
Here, we use the symbols $U$, $M$, and $b$ to represent the regularization matrix, as well as the matrix and vector introduced in Equation (\ref{source recon m}) and Equation (\ref{source recon b}), respectively.
The details of the calculation leading to the above result are shown in Appendix \ref{Sec dri strength}.

The amplitude-marginalized posterior is used for ranking and making inferences for lens models. Because the matrix $M$ and vector $b$ still depend on the lens parameters, this marginalized posterior is a function of the lens parameters.  We have omitted the lens parameters in the variable slot of the posterior function earlier for simplicity because the lens parameters are fixed for the derivation of the source reconstruction.

The source-marginalized posterior introduced above also provides a method for determining the source regularization strength, $\lambda_s$. If there is no strong prior on the regularization strength, it is advisable to use the regularization strength $\hat{\lambda}_s$ that maximizes the amplitude-marginalized posterior. When a flat prior for the strength is assumed, the recommended $\hat{\lambda}_s$ is obtained by solving the non-linear equation:
\begin{equation}
\begin{split}
\label{optimal lambda equation flat prior}
N_s = \hat{\lambda}_s &{\rm{tr}}\Big[(M+\hat{\lambda}_sU)^{-1}U\Big]
\\&+ \hat{\lambda}_s b^T (M+\hat{\lambda}_sU)^{-1}U(M+\hat{\lambda}_sU)^{-1}b,
\end{split}
\end{equation}
where $N_s$ is the dimension of the source amplitude $\{a_i\}$ vector space. A detailed and complete derivation of the optimal strength equation in our notation for natural weighting interferometric source reconstruction is provided in Appendix \ref{Sec dri strength}. This derivation of the optimal regularization strength follows \cite{mackay1992bayesian} and \cite{suyu2006bayesian} by exploiting Bayes' relations.

\section{Likelihood characterization using simulations}
\label{sec test likelihood}
In this section, we use simulated ALMA images to test the statistics of the likelihood function given by Equation (\ref{chi2 simplified}). We generate mock ALMA observations using \texttt{Lenstronomy} for the lensed sky model and \texttt{CASA} to convert the sky model into interferometric images. We then fit the simulated images using our likelihood function to characterize the performance.

\subsection{Simulation and Test Descriptions}
\label{sec simulation describe}
To simulate an ALMA observation, we first create the sky model image using \texttt{Lenstronomy}. We use parametric functions to model the sky. Specifically, we employ a Singular Isothermal Elliptical (SIE) lens density profile with external shear to simulate the mass distribution of the foreground lens galaxy, and a S\'{e}rsic elliptical light profile to serve as the light of the background source galaxy. The light of the lens galaxy is not included in the simulation, as lens galaxies are not bright enough at submillimeter wavelengths to contribute appreciably to the total flux. The definitions of the parametric profiles used in this paper are provided in Appendix \ref{Sec param trans}. For the parameter values used in the simulation, we adopt the best-fit parameters from the continuum data of the strongly-lensed galaxy SPT0346-52, the fitting of which will be discussed in Section \ref{sec fit 0346}.

\begin{figure*}[h!tp]
\centering
  \epsscale{0.9}
  \plotone{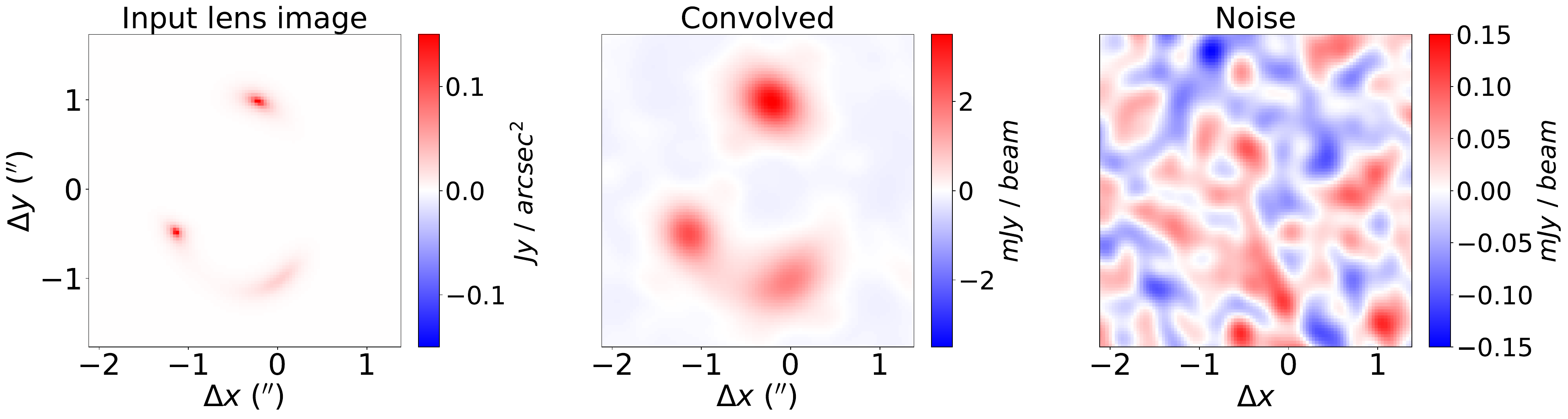}
  \caption{The model images and simulated ALMA noise. Left panel: The input model image generated by \texttt{lenstronomy}. Middle panel: The model image convolved with the dirty beam, generated by \texttt{CASA} and plotted in natural weighting. Right panel: One realization of ALMA random noise generated by \texttt{CASA}, plotted in natural weighting. The coordinates in all panels are $\Delta x$ and $\Delta y$ relative to the phase center of the simulation, where $\Delta x$ is in the \emph{negative} right ascension (RA) direction and $\Delta y$ is in the positive declination (DEC) direction.}
  \label{fig simulation with noise realization}
\end{figure*}

The input parameters for the SIE lens simulation are: $\theta_E=1.00''$, $e_{1,\rm{Lens}}=0.355$, $e_{2,\rm{Lens}}=-0.262$, $x_L=-0.598''$ and $y_L=-0.330''$. For the lens external shear, the parameters are: $\gamma_1=-0.012$ and $\gamma_2=-0.121$. For the background S\'{e}isic source, the paramters are: $I_0=0.023Jy$, $R_{\rm{Sersic}}=0.0944''$, $n_{\rm{Sersic}}=1.54$, $e_{1,\rm{source}}=-0.032$, $e_{2,\rm{source}}=0.111$, $x_s=-0.420''$, and $y_s=-0.095''$. The angular position parameters are defined relative to the ALMA phase center. The model image generated with this set of parameters is shown in the left panel of Figure \ref{fig simulation with noise realization}.

We then use \texttt{CASA} to make mock interferometric observations of the sky model. The \texttt{simobserve} tool is used to simulate ALMA observations. The antenna configurations are extracted from the weblog of SPT0346-52 (project 2013.1.00880.S), and the total observation time is set to $2751$ seconds to mimic the real SPT0346 observation. This simulation setup results in a naturally weighted synthesized beam with a resolution of approximately $0.38''$. The naturally weighted dirty image from this simulation is shown in the middle panel of Figure \ref{fig simulation with noise realization}. To simulate the noise, we use the \texttt{simulator.setnoise} tool in \texttt{simplenoise} mode to add Gaussian noise to the real and imaginary parts of the visibility data on the $uv$-plane. We use random seeds of the \texttt{simulator} tool to generate an ensemble of ten realizations of noise, producing ten differently noisy ALMA measurements of the same input sky model. One of the ten noise realizations is shown in the right panel of Figure \ref{fig simulation with noise realization} in natural weighting. The simulated noise in natural weighting has an overall rms of $36.5\mathrm{\mu Jy/beam}$, and this noise $\sigma$ is the noise level input to the likelihood function given by Equation (\ref{chi2 simplified}). Finally, we use \texttt{tclean} to generate the dirty image, dirty beam (PSF), and primary beam (PB) for the ten realizations. The dirty beam and primary beam are the same for all ten realizations in this simulation. We plot the dirty images in the pixel size of $0.035''\times 0.035''$, which is much smaller than the synthesized beam size of this simulation. Additionally, the $0.035''$ pixel length has a Nyquist frequency that exceeds the spatial frequency corresponding to the longest baseline of this fake observation, ensuring that the images retain the full information of the data.

To study the behavior of the image-plane likelihood function, we perform an initial test to fit the simulated ensemble of ALMA data using our image-plane method and compare the results with the fitting of the same simulated data using a $uv$-plane lens modeling code, \texttt{visilens} \citep{hezaveh2013alma,spilker2016alma}. The result of the first test shows that the behavior of the MCMC chain in our image-plane fitting is similar to that in the $uv$-plane fitting. It also demonstrates that the proportions of the confidence intervals predicting the correct input values are comparable to the percentiles of the intervals, which justifies the likelihood functions for both the image-plane and $uv$-plane fittings. We then perform a second test to examine the bias from the result of test 1. The result of the second test shows that the bias introduced by the flat sampling priors becomes smaller as the signal-to-noise ratio increases.

\begin{figure*}[h!tp]
\centering
  \epsscale{1}
  \plotone{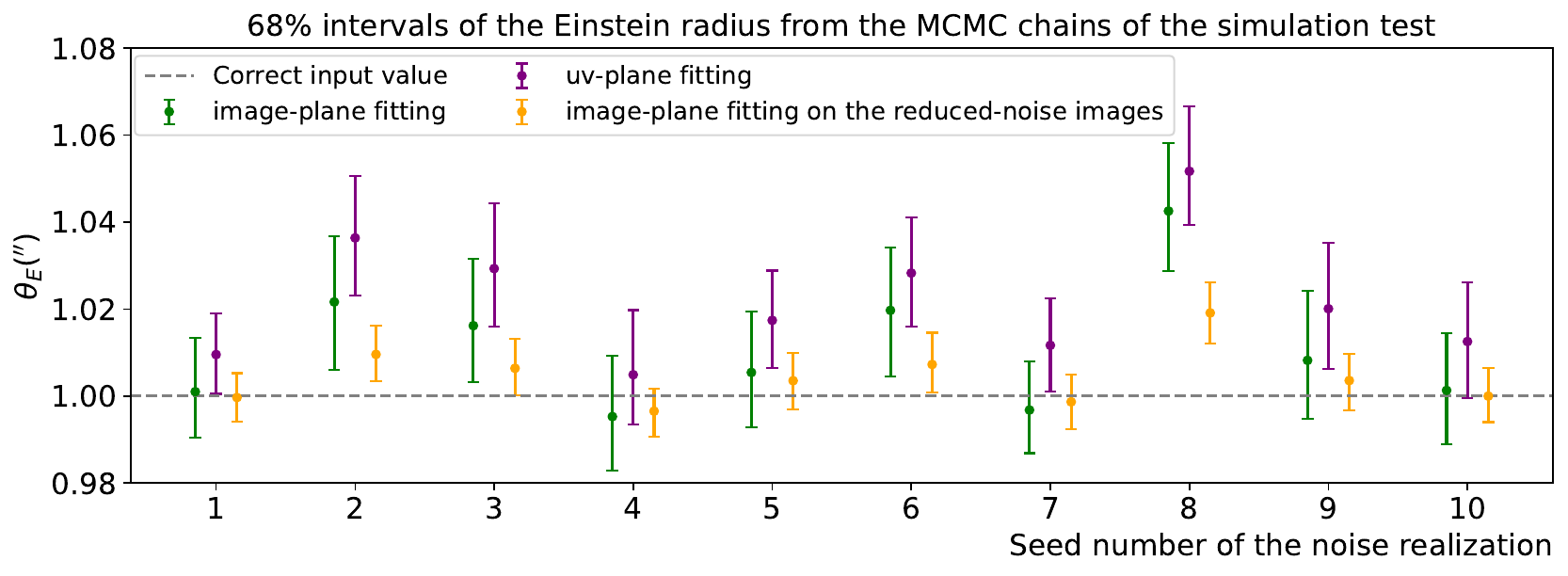}
  \caption{$1\sigma$ (68\%) intervals of the Einstein radius from the MCMC chains of the fittings on the simulated images. The green and purple vertical bars correspond to the fittings using the image-plane method and the $uv$-plane method, respectively (test 1, Section \ref{sec simulation test 1}), while the orange bars represent the fittings on the reduced-noise images from test 2 (Section \ref{sec simulation test 2}). Each vertical bar spans from the 16th to the 84th percentile, with the middle dot indicating the median (50th percentile). The horizontal axis shows the seed number of the noise. The dashed horizontal line denotes the correct input value of the Einstein radius, and its intersections with the vertical error bars indicate the intervals that contain the correct input value. }
  \label{fig mcmc 1 sigma thetaE}
\end{figure*}

\begin{figure}[htp]
  \epsscale{1.15}
  \plotone{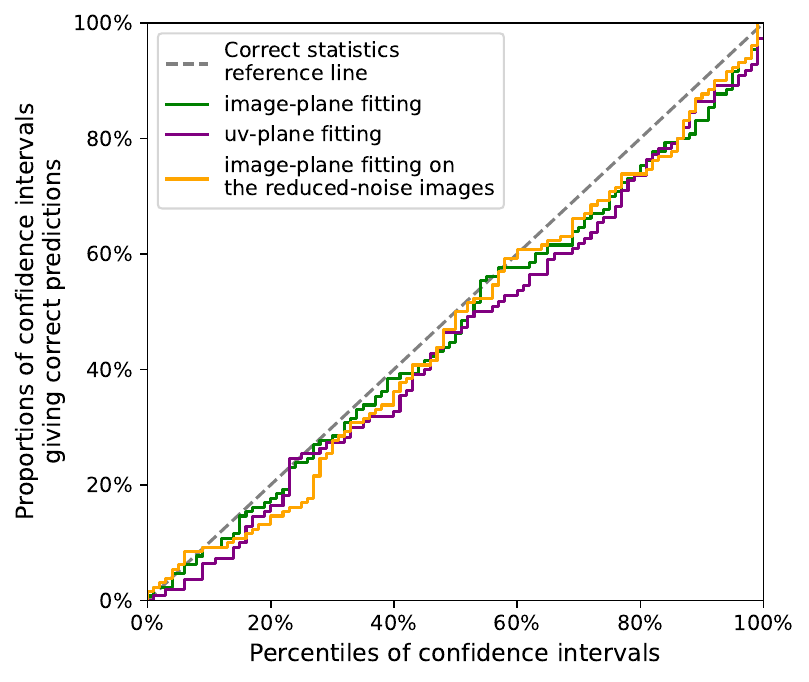}
  \caption{Plot of the proportions of confidence intervals that contain the true input values versus the confidence interval percentiles. For the image-plane fittings, 13 parameters across 10 realizations are included, resulting in 130 total parameter estimates. For the $uv$-plane fittings, 110 total parameter estimates are included, excluding the lens position parameters.
  The horizontal axis represents the percentile, while the vertical axis shows the proportions. The gray dashed line serves as the ideal statistical reference. }
  \label{fig mcmc error bar ratio}
\end{figure}

\subsection{Tests and Results}
\subsubsection{Test 1: Compare the Fitting Results of the Image-plane and $uv$-plane Fitting Techniques}
\label{sec simulation test 1}

We perform lens modeling of the simulated data using our image-plane approach. For comparison, we also fit the data using the $uv$-plane fitting code.

For the image-plane fitting, we fit the simulated naturally weighted dirty images of each realization separately. We perform the fitting using \texttt{Lenstronomy}, equipped with our image-plane likelihood function. During the \texttt{Lenstronomy} fitting, we fix the external shear center to the ALMA image phase center and allow other parameters to vary within large intervals that include the correct input values. After completing the MCMC sampling, we convert the source center positions into source offset positions using Equation (\ref{source offset position}) for each element of the MCMC chains to ensure consistency with the parameters defined in \texttt{visilens}.

We extract the $68\%$ percentile ($1\sigma$) intervals from the MCMC chains for each parameter and each image realization. We plot the $1\sigma$ intervals of the lens Einstein radius in Figure \ref{fig mcmc 1 sigma thetaE} as a representative example to discuss the fitting results. The $1\sigma$ confidence intervals of other parameters exhibit similar statistical behavior to those of the Einstein radius and are plotted in Figure \ref{fig mcmc 1 sigma others} (in Appendix \ref{sec appendix add plots}).

The $1\sigma$ intervals are distributed close to the correct input value. However, as shown in Figure \ref{fig mcmc 1 sigma thetaE}, the average of the predicted parameter is larger than the correct input value, suggesting a positive bias in the estimation of the Einstein radius. We will address this bias later in Section \ref{sec simulation test 2}.

Quantitatively, approximately $62\%$ of the total $1\sigma$ intervals for all parameters contain the corresponding correct input values. While ideal statistical expectations suggest that the probability of the $1\sigma$ intervals containing the correct value should be $68\%$, the $62\%$ ratio from our image-plane fitting is reasonable. Furthermore, we extract confidence intervals covering the full range from $0\%$ to $100\%$. For each parameter and each realization, we start from the medians of the MCMC chains. We then plot the percentiles against the proportions of intervals that contain the correct input values in Figure \ref{fig mcmc error bar ratio}. As Figure \ref{fig mcmc error bar ratio} shows, the proportions of intervals containing the correct input values are close to the corresponding percentiles, suggesting that the natural weighting likelihood function of our method provides an acceptable statistical description of the random visibility noise added by the ALMA simulation.

Next, we fit the simulated data using \texttt{visilens}, which also incorporates MCMC sampling. Since the free parameters used to describe the SIE and S\'{e}rsic profiles differ between \texttt{visilens} and \texttt{lenstronomy}, we transform the \texttt{visilens} parameters into the free parameters used by \texttt{lenstronomy} after fitting. Due to the different conventions for the image-plane coordinate zero point in \texttt{visilens} and \texttt{Lenstronomy}, the best-fit lens center coordinates from \texttt{visilens} do not match the correct input values from \texttt{Lenstronomy}. Therefore, the quantitative analysis of the $uv$-plane fitting does not include the lens center parameters.

As with the image-plane fitting, we also plot the $1\sigma$ intervals for the $uv$-plane fitting in Figure \ref{fig mcmc 1 sigma thetaE} and Figure \ref{fig mcmc 1 sigma others} (in Appendix \ref{sec appendix add plots}). The plots show that the $1\sigma$ error bars of the $uv$-plane fitting are distributed around the correct input values and are comparable in length to those of the image-plane fitting, confirming that the results of the image-plane fitting are as reliable as those of the $uv$-plane fitting. We also plot the proportions of intervals containing the correct input values for the $uv$-plane fitting in Figure \ref{fig mcmc error bar ratio}. As the plot shows, the $uv$-plane fitting functions correctly, further validating both the image-plane and $uv$-plane approaches for lens modeling.

While the above results demonstrate that the image-plane fitting performs as well as the $uv$-plane fitting, another observation can be made regarding the distribution of the $1\sigma$ intervals: the distribution appears biased. Taking the $1\sigma$ intervals of the Einstein radius (Figure \ref{fig mcmc 1 sigma thetaE}) as an example, we find that the error bars are more likely to be distributed above the correct input value than below it. However, since this occurs in both the image-plane and $uv$-plane fittings, the uneven distribution of confidence intervals is not caused by the image-plane likelihood method. Instead, we suggest that this bias is introduced by the flat sampling priors used for the parameters during the parametric fitting, as briefly mentioned in Section \ref{sec para fitting flat prior}. In test 2 below, we further explore this idea.

\subsubsection{Test 2: Fitting the Reduced-noise Images to Understand the Sampling Prior Bias}
\label{sec simulation test 2}

To see that the bias in the posterior distributions is introduced by the sampling priors, we adjust the relative weights of the likelihood and prior functions within the posterior.

We increase the weight of the likelihood function from test 1. To do this, we reduce the noise level of the 10 simulated images by half during the simulation and keep the noise pattern unchanged to create the simulated images. When the noise level is halved, the peak SNR of the simulated images increases. Ideally, if the noise level decreases, the denominator of the $\chi^2$ becomes smaller. As a result, the likelihood function becomes sharper at its maximum, leading to shorter error bars and a smaller bias in the posterior.

To verify this idea, we use the image-plane likelihood function to fit the set of reduced-noise images. After the MCMC is completed, we plot the $1\sigma$ error bars of the parameters in Figure \ref{fig mcmc 1 sigma thetaE} and Figure \ref{fig mcmc 1 sigma others}. As the plots show, the $1\sigma$ error bars of test 2 shrink toward the correct input values. This result meets our expectation that a sharper likelihood at the maximum reduces the sampling prior bias.

We also plot the proportions of the confidence intervals providing correct predictions versus the percentiles of the intervals in Figure \ref{fig mcmc error bar ratio}. The plot shows that the fitting of the reduced-noise images is in overall agreement with the correct statistics.

\section{Fitting Results of ALMA Data}
\label{sec fitting alma data}
Having tested our likelihood function on the simulated data, we further apply our method to two well-studied strongly lensed sources observed with ALMA: SPT0346-52 and SPT0311-58.

\subsection{Parametric Fitting of SPT0346-52}
\label{sec fit 0346}

\begin{figure*}[htbp]
  \epsscale{1.1}
  \plotone{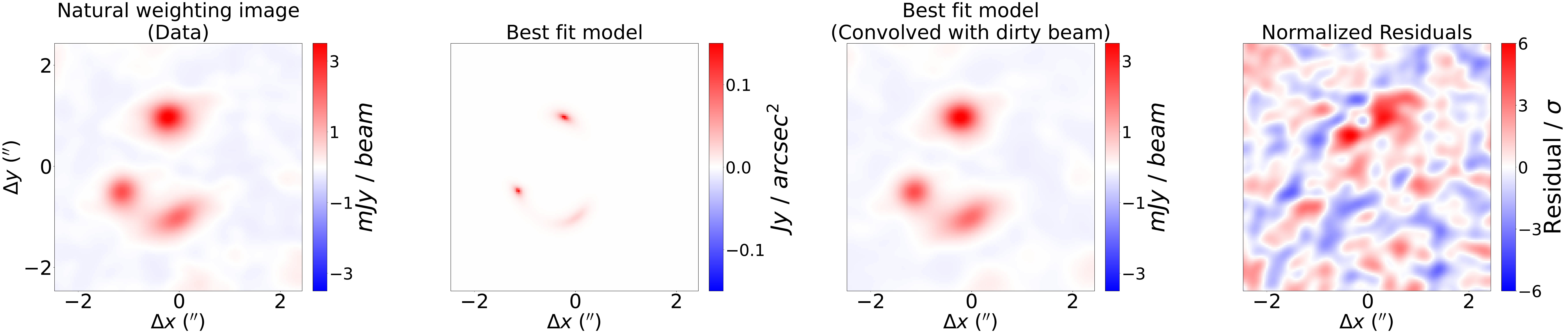}
  \caption{Natural weighting dirty image of SPT0346 and the parametric fitting result. All images have a pixel size of $0.035''$ and are $140\times140$ pixels in size. The horizontal and vertical coordinates are relative to the phase center of this image.
  Leftmost panel: naturally weighted image of SPT0346. Left-middle panel: best-fit parametric model. Right-middle panel: dirty image of the best-fit model. Rightmost panel: normalized residuals of the parametric fitting in the image plane.
  }
  \label{fig spt0346 fitting result image plane}
\end{figure*}

\begin{figure*}[ht]
  \epsscale{1.15}
  \plotone{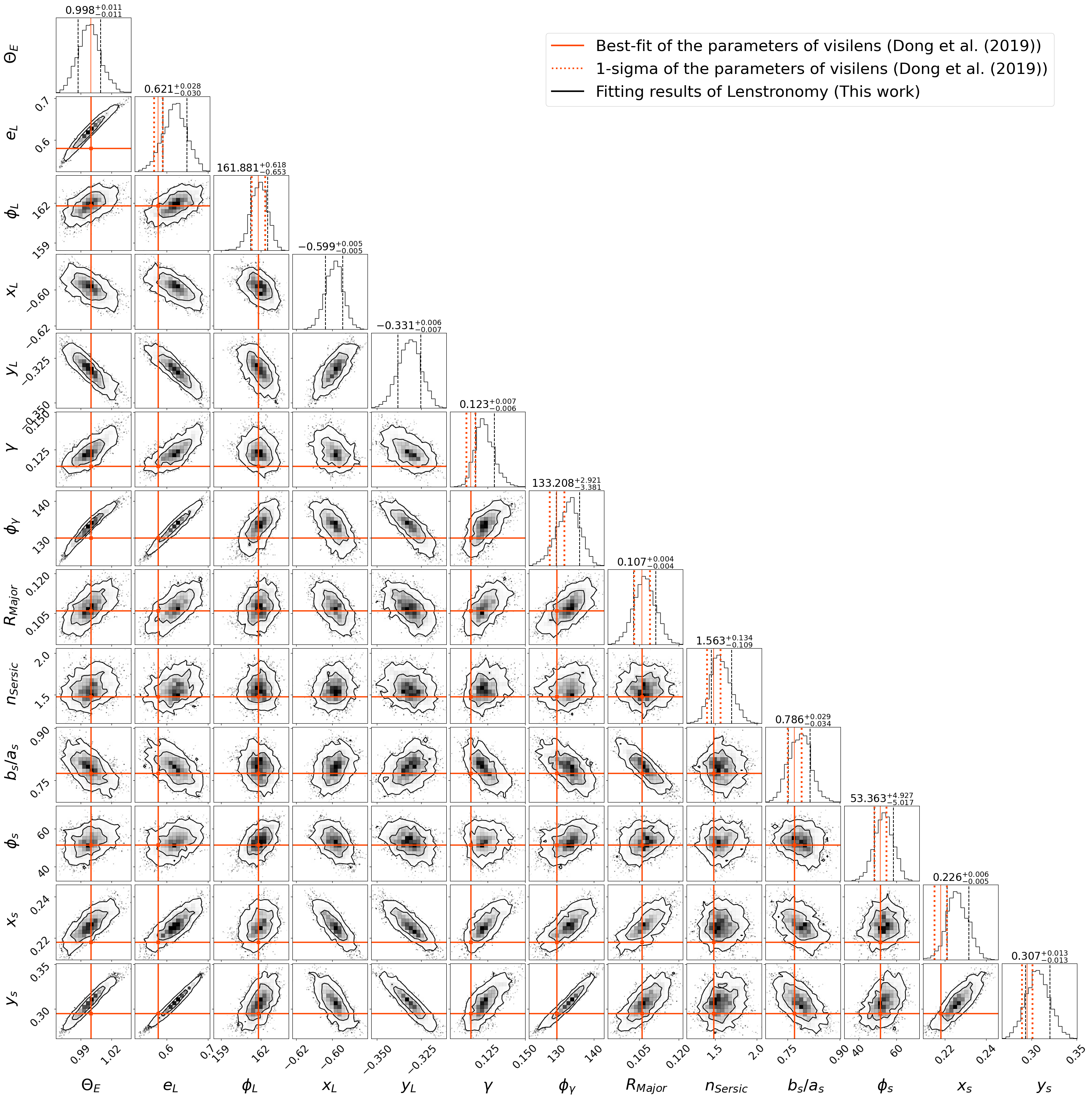}
  \caption{Corner plot of the MCMC fitting for SPT0346. In the 2D distributions, the black inner and outer contours represent the $1\sigma\ (68\%)$ and $2\sigma\ (95\%)$ confidence regions. In the 1D distributions, the black dashed lines indicate the $1\sigma$ interval for each parameter, and the titles of each 1D plot display the corresponding parameter's $1\sigma$ interval numerically. The red solid lines show the best-fit values from \cite{dong2019source}, obtained by fitting the same data using \emph{visilens}. In the 1D marginalized distributions, the dashed red lines mark the $1\sigma$ interval from the reference. The dashed red lines for the Einstein radius are absent because the reference uncertainty for this parameter is $0.10''$, which exceeds the scale of this plot. The $68\%$ confidence intervals for the lens positions $(x_L, y_L)$ differ between our fit and the reference due to differences in coordinate conventions. The corner plot is generated using the code from \cite{corner}.}
  \label{fig spt0346 corner}
\end{figure*}

We use the continuum observations of SPT0346-52 to test the parametric fitting. The specific data for the SPT0346-52 continuum observations come from ALMA Cycle 2 project 2013.1.00880.S. The details of the observations are described in \cite{dong2019source}. In the analysis below, we refer to the SPT0346-52 strong lensing system simply as SPT0346.

We use CASA \texttt{tclean} to process the visibility data and produce the naturally weighted dirty image, dirty beam, and primary beam. We set the pixel length and width to $0.035^{''}$ and generate images with a size of $480\times480$ pixels. This naturally weighted dirty image yields a beam size of approximately $0.47^{''}\times0.39^{''}$. Therefore, the pixel size of our image is smaller than the beam size, and the Nyquist frequency corresponding to this pixel size is well beyond the frequency of the longest baseline. To save computation time, we crop the image to $140\times140$ pixels. The smaller image still contains the majority of the flux from the SPT0346 system, as shown in the leftmost panel of Figure \ref{fig spt0346 fitting result image plane}.

We then fit the SPT0346 image parametrically using our code, as described in Section \ref{sec methodology}. Similar to \cite{dong2019source} and the simulation work presented above, we model the lens density profile using a Singular Isothermal Ellipsoid (SIE) with external shear and model the source surface brightness with an elliptical S\'{e}rsic light profile.

We need to estimate the noise level of the dirty image for likelihood computation. We estimate the noise level by fitting the dirty image twice. First, we make an initial estimation of the noise $\sigma_n$ by taking the rms of the region of the dirty image that is away from the flux of the lensing system. This initial estimation is not accurate, as it includes part of the source flux introduced by the dirty beam sidelobes. We perform an initial fit using this initial $\sigma_n$, and after the fitting, we obtain a residual map that is close to the noise map. We then take the rms of the pixels in the residual map as our final estimation of the noise level, yielding $\sigma_n = 30.8\mathrm{\mu Jy/beam}$. The noise level estimated in this way is close to the noise level ($\sigma_n = 32.5 \mathrm{\mu Jy/beam}$) reported by \cite{dong2019source}. We then fit the image again using our newly estimated $\sigma_n$. With the improved estimation of $\sigma_n$, the MCMC sampling provides more accurate uncertainties for the parameters.

Alternatively, we can take advantage of the CLEAN algorithm to estimate the noise level. Specifically, after CLEAN is performed on the dirty image, a residual map of the dirty image is produced, and we take the rms of the residual map to obtain $\sigma_n$. While we do not use that method in this paper, we have verified on simulated data used in Section \ref{sec simulation test 1} that this method returns noise estimates that are typically accurate to within $5\%$, which is similar to the accuracy of the method outlined above.

The best-fit model and the corresponding image-plane residuals are shown in Figure \ref{fig spt0346 fitting result image plane}. From the residual map, we see that the residuals of the bottom two arcs are low. Quantitatively, the residuals in the lower half ($\Delta y<0$) are entirely within $4\sigma$. However, there are large residuals around the upper flux region, where the highest residual level reaches $6.1\sigma$. The residual map provided by \cite{dong2019source} also shows a high-residual region at the same location, where the residuals exceed $4\sigma$. Therefore, the common residual behavior observed in both our fitting result and that of \cite{dong2019source} suggests that the high-residual region is due to the limited flexibility of the analytical lens and source profiles.

We show the MCMC corner plot of our fitting in Figure \ref{fig spt0346 corner}. In the plot, we have also included the best-fit values and $1\sigma$ error bars from the analysis of \cite{dong2019source} for comparison. We find that the $1\sigma$ error bars from our image-plane fitting method overlap with the $1\sigma$ intervals from the reference for nearly all parameters, except for the lens center positions, for which the offset is relatively large. Quantitatively, the offsets of $x_L$ and $y_L$ are approximately $0.031^{''}$ and $0.038^{''}$ respectively, which are close to the image pixel size ($0.035^{''}$). This suggests that the offset of the lens center is due to differences in the definitions of the zero coordinate and the pixelation of the image grids between the two methods, as discussed in Section \ref{sec simulation test 1}.

Since this is a comparison of fittings to the same data, in principle the image-plane and $uv$-plane fitting methods should agree perfectly. The parameter offsets observed here (and in the simulations of Section \ref{sec simulation test 1}) indicate the presence of systematic errors in one or both of the methods. However, these systematic errors are smaller than the statistical uncertainties.

\subsection{Source Reconstruction of SPT0311-58}
We perform pixelated source reconstruction on the dust continuum image of SPT0311-58, a complex galaxy system at high redshift, using the image-plane likelihood function. The ALMA observations of SPT0311-58 analyzed in this work were taken from projects 2016.1.01293.S and 2017.1.01423.S. The details of the observations and the calibration procedure are described in \cite{spilker2022chaotic}. In the analysis below, we refer to the strong lensing system SPT0311-58 as SPT0311.

The SPT0311 images are generated using \texttt{CASA} \texttt{tclean}. Specifically, we produce the pixelated natural weighting dirty image, dirty beam, and primary beam from the calibrated visibilities. The image is created with a pixel size of $0.008^{''}$. We work with an image cut of $450\times450$ pixels, which contains the majority of the flux from the lensed SPT0311 galaxies. This dirty image cut is shown in the lower leftmost subplot of Figure \ref{fig spt0311 recon demo}.

Unlike SPT0346, SPT0311 has a complex source structure, which is difficult to describe using a few S\'{e}rsic profiles. A more flexible source model, such as a set of source pixels, is therefore needed to fit the source flux distribution. In this work, we perform the pixelated source reconstruction of SPT0311 using the method based on the image-plane likelihood described in Section \ref{sec source recon equations}. We then compare our reconstruction result with the previous analysis in \cite{spilker2022chaotic}.

\subsubsection{Reconstructing the East and West Galaxies Separately}
\label{sec source recon procedure}

\begin{figure*}[ht]
\epsscale{1.0}
\plotone{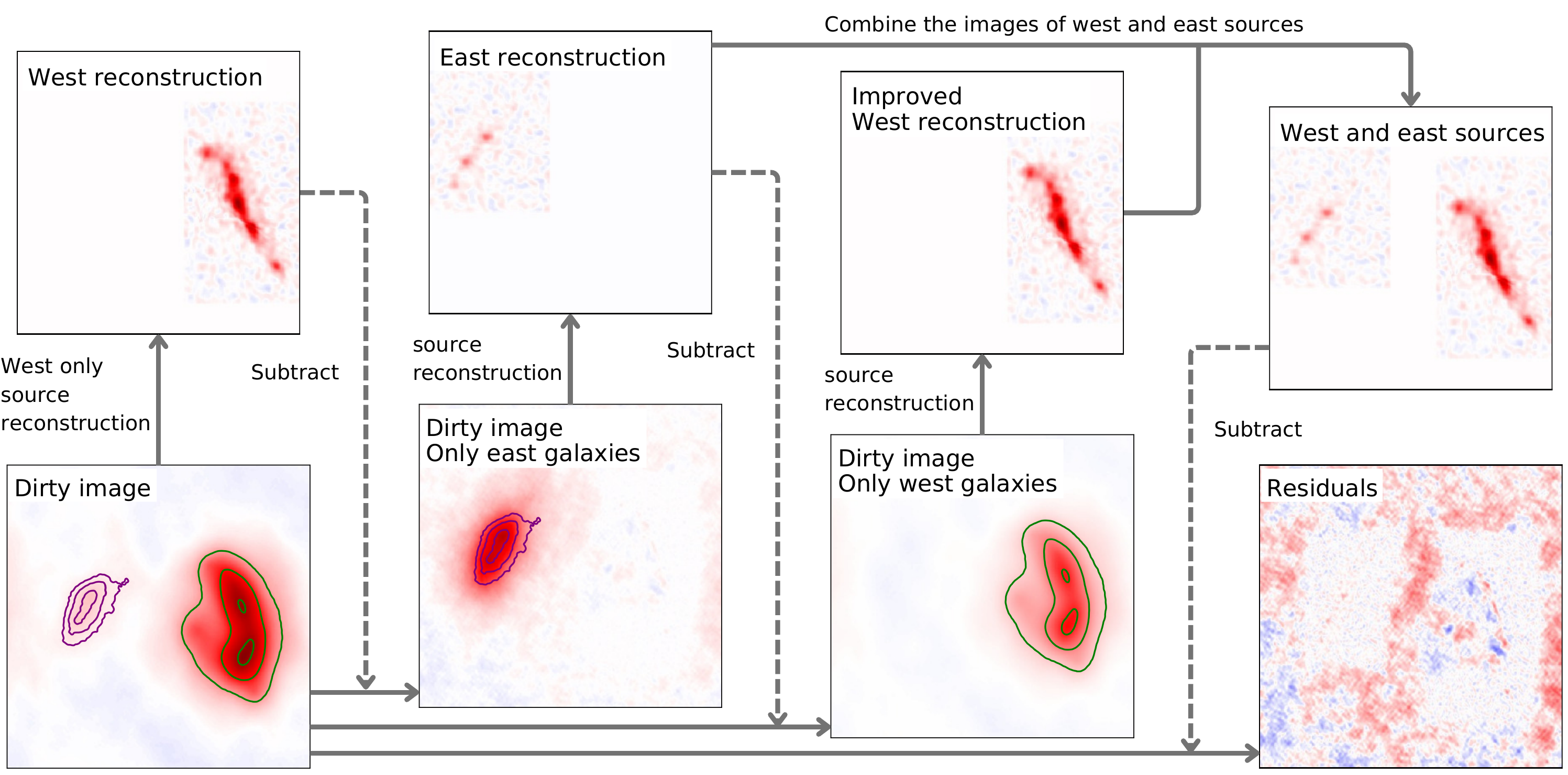}
\caption{Demonstration of the separate source reconstruction of SPT0311. Note that the color scales vary among the dirty images, as they are chosen to best display the flux in each subplot.}
\label{fig spt0311 recon demo}
\end{figure*}

SPT0311 consists of two sets of lensed sources: the east galaxies and the west galaxies. The dirty image shown in the lower leftmost subplot of Figure \ref{fig spt0311 recon demo} shows a gap region between the east and west detections. Based on prior knowledge of the SPT0311 strong lensing system from previous analysis \citep{spilker2022chaotic}, the flux of the east and west detections originates from different galaxies. The west galaxies are strongly lensed by the foreground lensing galaxies. The east galaxies are shifted by the lensing effect but are not significantly magnified, as they are farther from the lensing center on the sky map. To save computational resources, we choose to reconstruct the west and east galaxies separately.

To reduce mutual influence between the two source regions, we reconstruct them independently. As Figure \ref{fig spt0311 recon demo} demonstrates, we first reconstruct the west sources from the dirty image. We then subtract the lensed model of the west sources, resulting in a dirty image that contains only the flux from the east galaxies. This approach makes the subsequent reconstruction of the east galaxies less affected by the presence of the west galaxies. Next, we apply the same process for west galaxies. We subtract the lensed model of the east galaxies from the original dirty image, obtaining a dirty image that contains only the west galaxies. We then perform an improved reconstruction of the west galaxies using this image. The improved reconstruction of the west galaxies is now less influenced by the east galaxies. Finally, we combine the reconstructed east galaxies and the improved west galaxies to obtain the final reconstruction of both the east and west galaxies.

For the reconstruction of the west galaxies, we choose a rectangular region of the source plane with a size of $120\times200$ pixels. In angular positional coordinates, this region covers $(-0.56'', 0.40'')$ in $\Delta x$ and $(-0.92'', 0.68'')$ in $\Delta y$, where $\Delta x$ and $\Delta y$ are measured relative to the ALMA phase center. For the east galaxies, we perform the reconstruction within a $120\times140$-pixel rectangular region of the source plane, corresponding to $(-1.88'', -0.92'')$ in $\Delta x$ and $(-0.36'', 0.76'')$ in $\Delta y$.

The pixelated source reconstruction follows the procedure described in the flow chart in Figure \ref{fig flow chart} (described in Section \ref{sec source recon equations}). We map source pixels from the source plane to the image plane using the lens equation with an SIE lens model. We then compute the matrix $M$ and vector $b$ as defined in Equation (\ref{source recon b}) and Equation (\ref{source recon m}). To avoid overfitting, we apply gradient regularization, with its matrix form introduced in Appendix \ref{sec appendix regularization}. We numerically determine the optimal regularization strength $\lambda_s$ by solving Equation (\ref{optimal lambda equation flat prior}). Finally, we solve for the most probable flux of each source pixel (i.e., the $\{a_i\}$) by solving the linear equations (Equation \ref{linear amp solving equation with prior}). The collection of the solved source pixel values forms the pixelated source image.

For consistency in the source prior across the entire source map, we use the same regularization strength $\lambda_s$ for both the east and west reconstructions. This value of $\lambda_s$ is determined during the first reconstruction (see Figure \ref{fig spt0311 recon demo}) of the west galaxies.

We now apply the reconstruction procedure to SPT0311 and discuss the results.

\begin{table*}[ht]
\centering
\begin{tabular}{c|ccccc}
\hline \hline
\begin{tabular}{c}
SIE Lens parameters\\
for SPT0311 reconstruction
\end{tabular}
& $\theta_E('')$ & $e_{1,L}$ & $e_{2,L}$& $x_L('')$& $y_L('')$ \\
\hline
Caustics Calibration & 0.330 & 0.323 & -0.322 & -0.375 & -0.136 \\
Parametric Fitting & $0.331^{+0.001}_{-0.002}$ & $0.334^{+0.005}_{-0.005}$ & $-0.446^{+0.010}_{-0.011}$ & $-0.373^{+0.001}_{-0.001}$ & $-0.139^{+0.001}_{-0.001}$ \\
\hline
\end{tabular}
\caption{SIE lens parameters for the SPT0311 reconstruction. Two SIE lens models are listed here. The $x_L$ and $y_L$ denote the center coordinates of the SIE lens relative to the phase center.}
\label{tab spt0311 fit parameters}
\end{table*}

\subsubsection{Source Reconstruction with the ``Caustics Calibration'' Lens Model}

First, we choose to perform the reconstruction using a lens model similar to that in \cite{spilker2022chaotic}, to facilitate comparison with their results. However, since the direct transformation of lens parameters from the package used in \cite{spilker2022chaotic} is not well calibrated to the lens parameters defined in \texttt{Lenstronomy}, we instead adopt a lens model that matches the caustics of the SIE lens used in the reference.

\begin{figure*}[!tp]
  \epsscale{1.16}
  \plotone{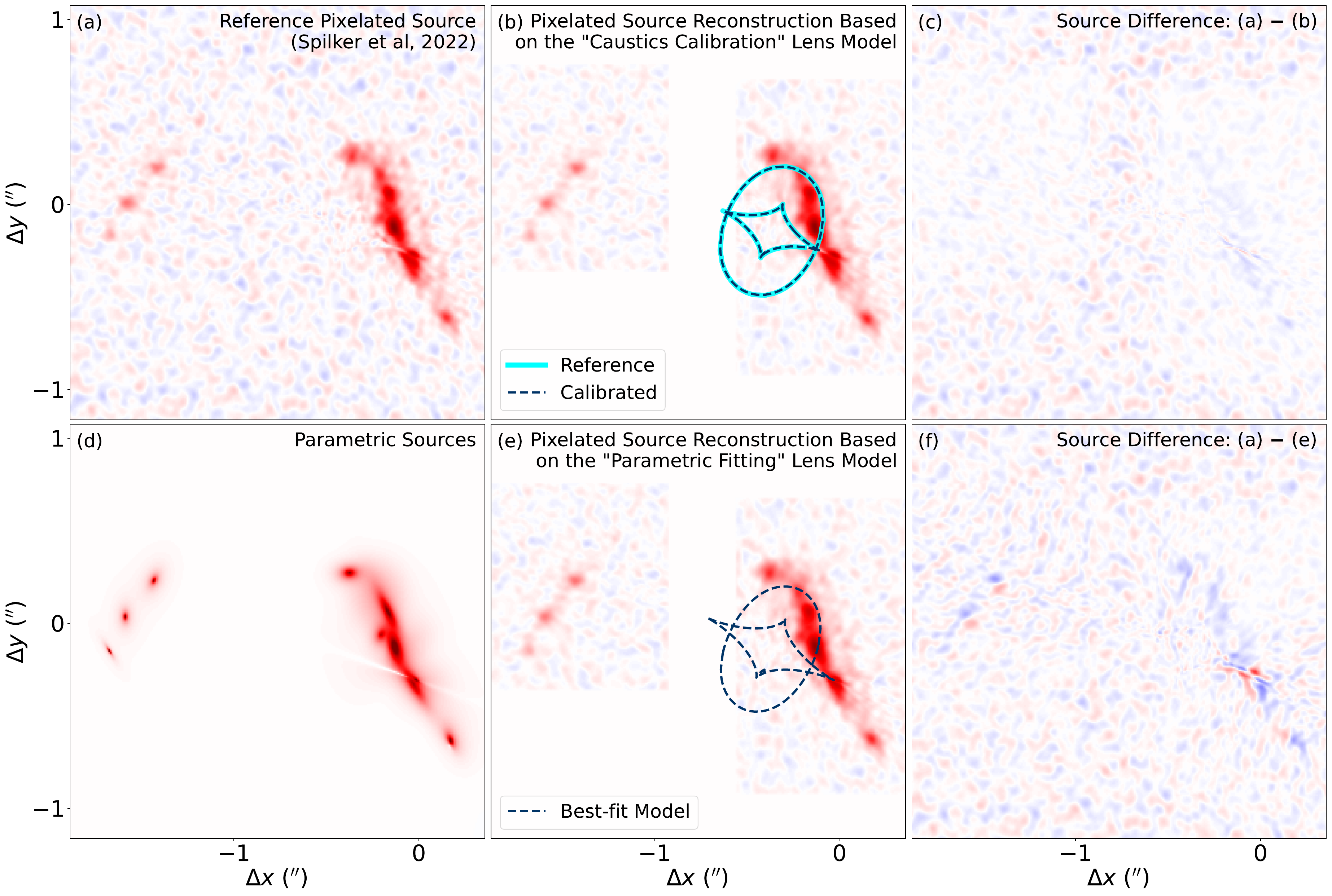}
  \caption{Source reconstruction maps and comparison with the reference \citep{spilker2022chaotic}. Each subplot is labeled in its upper-left corner, and all subplots use the same color scale. The coordinates in each plot are relative to the phase center. Subplot (a) shows the reference source reconstruction. Subplots (b) and (e) show the image-plane-based pixelated source reconstruction from the ``Caustics Calibration'' lens model and the ``Parametric Fitting'' lens model, respectively. Subplots (c) and (f) show the differences from the reference reconstruction. Subplot (d) shows the parametric source of the best-fit parametric model.}
  \label{fig spt0311 recon sources}
\end{figure*}

To match the lens model, we adjust five parameters of the SIE lens: the Einstein radius ($\theta_E$), the ellipticity components ($e_{1,L}, e_{2,L}$), and the lens center position ($x_L, y_L$). Among these parameters, the definition of ellipticity is relatively unambiguous. Thus, we fix the ellipticities $e_{1,L}$ and $e_{2,L}$ to the values $(0.323, -0.322)$, obtained from the direct transformation of the SIE lens parameters used in \cite{spilker2022chaotic}. We then adjust $\theta_E$, $x_L$, and $y_L$ to match the caustics of our SIE lens model with those plotted in \cite{spilker2022chaotic}. The SIE lens model obtained through this process is referred to as ``Caustics Calibration'' in Table \ref{tab spt0311 fit parameters}. The overlap between the matched caustics and the reference caustics is shown in subplot (b) of Figure \ref{fig spt0311 recon sources}.

Using the ``Caustics Calibration'' lens model, we perform source reconstruction for the west and east galaxies following the procedure described in Section \ref{sec source recon procedure}. The reconstructed source is shown in subplot (b) of Figure \ref{fig spt0311 recon sources}. For comparison with the source reconstruction in \cite{spilker2022chaotic}, we also include their reconstructed source in subplot (a). Additionally, subplot (c) presents the difference between our reconstructed source and theirs.

The source difference map, shown in subplot (c) of Figure \ref{fig spt0311 recon sources}, indicates that the difference is generally small. The pixel values in the difference map remain within $\pm1.1\mathrm{\mu Jy/pixel}$. This small difference confirms that our image-plane-based reconstruction approach produces results highly consistent with those of \cite{spilker2022chaotic}, when the lens model is matched.

The ``Caustics Calibration'' lens model could be further improved, as the ellipticities were fixed and not optimized in the parameter space during the caustic calibration process.

\subsubsection{Parametric Fitting and the Source Reconstruction with the ``Parametric Fitting'' Lens Model}
\label{sec spt0311 parametric fitting}
Using parametric models to fit the strong lensing system of SPT0311 is challenging because the source structure is complex and difficult to describe accurately with parametric sources. To account for this complexity, we fit the system using a large number of S\'{e}rsic sources. Specifically, we use three S\'{e}rsic profiles to fit the three brightest sources of the east galaxies and up to ten profiles for the west galaxies. To further capture the irregular and faint extended flux from the west background galaxies, we also include a set of shapelets in the source model. The shapelet source profile is defined in \cite{refregier2003shapelets}, and examples of parametric fitting with a combination of S\'{e}rsic sources and a shapelet background can be found in \cite{shajib2019every} and \cite{birrer2019h0licow}.

To perform the parametric fitting of SPT0311, we adopt the ``Caustics Calibration'' lens model as the initial guess for the lens parameters. Starting from this model, we gradually increase the complexity of the source model. We place three S\'{e}rsic profiles around the centers of the three brightest east galaxies and begin with three S\'{e}rsic profiles for the west galaxies. After the fitting converges, we add one more S\'{e}rsic source for the west galaxies and repeat the fitting process. We iterate this process until there are ten S\'{e}rsic source profiles for the west galaxies. We then add a shapelet set centered on the west galaxies. The resulting best-fit parametric source map is shown in subplot (d) of Figure \ref{fig spt0311 recon sources}. The corresponding best-fit lens model parameters are listed as the ``Parametric Fitting'' lens model in Table \ref{tab spt0311 fit parameters}.

We make two clarifications regarding this parametric fitting. First, stopping the fitting at ten S\'{e}rsic light profiles for the west galaxies does not imply that the parametric fitting cannot be made more complex. In principle, the rejection or adoption of a more complex model should be determined based on statistical hypothesis testing. However, due to limited computational resources, we do not pursue additional complexity once we achieve residuals consistent with the noise level. Second, using ten S\'{e}rsic profiles (and a shapelet set) to model the west galaxies does not imply that there are ten individual galaxies. As shown in subplot (d), the best fit is achieved by combining multiple source profiles rather than assigning one profile per galaxy. A detailed scientific analysis of the west galaxies is discussed in \citep{spilker2022chaotic} and is beyond the scope of this work.

Using the ``Parametric Fitting'' lens model, we perform source reconstruction and plot the pixelated source in subplot (e) and the difference from the reference source in subplot (f) of Figure \ref{fig spt0311 recon sources}. The lens model obtained through this approach is more elliptical than the ``Caustics Calibration'' lens model, as indicated by their parameters in Table \ref{tab spt0311 fit parameters}. As a result, the reconstructed source is distributed slightly differently. The difference is most apparent in the lower region of the west galaxies, as shown in the difference map in subplot (f). However, the difference is small compared to the overall source fluxes. This confirms that the source reconstruction obtained from the ``Parametric Fitting'' approach has reproduced similar features of the background source light distribution.

In principle, parametric fitting followed by source reconstruction can be used to analyze complex background galaxies, as in the case of SPT0311. However, this is not always the optimal approach. As parametric models become more complex and the number of degrees of freedom increases, it becomes increasingly difficult for fitting algorithms such as PSO or MCMC to find the maximum of the posterior function in the high-dimensional parameter space. To address this numerical challenge, source reconstruction techniques at each iteration are preferred. The amplitudes of the source profiles can be inferred linearly, keeping the number of non-linear parameters small. The semi-linear inference technique \citep{warren2003semilinear} performs fitting with parametric lens models while reconstructing the source at each iteration. This approach is useful for analyzing strongly lensed sources with complex morphology, as demonstrated in the study of SPT0311 by \cite{spilker2022chaotic}. We plan to improve our code to incorporate this functionality for interferometric images in future work.

\section{Conclusions}
\label{Sec discussion}
In this work, we presented an image-plane likelihood function for analyzing interferometric images. This approach benefits from the proportionality between the noise covariance matrix and the dirty beam under natural weighting. We conducted the first-ever test of the image-plane likelihood function in lens modeling, performing fittings on simulated ALMA observations as well as real ALMA data images and comparing the results with the $uv$-plane lens modeling results published in previous works.

Our test on the simulated images shows that fittings with the image-plane likelihood function provide correct uncertainty estimations for the lens and source parameters. Specifically, the parameter confidence intervals from the ensemble of fittings are consistent with their statistical definitions. Our test also verifies that fittings using the $uv$-plane approach yield expected statistical properties in parameter predictions. The difference in the best-fit lens center parameters between the image-plane and $uv$-plane approaches can be explained by the different astrometric coordinate conventions used in the two codes. After accounting for the difference in astrometric conventions, the parametric fitting and source reconstruction of the real ALMA images using the image-plane approach are generally consistent with previous analyses using the $uv$-plane approach. These test results demonstrate that the lens modeling techniques based on the image-plane likelihood function, as presented in this work, are reliable and capable of producing results comparable to the traditional $uv$-plane approach.

The positive test results of the image-plane lens modeling technique for interferometric images suggest that this method is effective for future studies of strong lensing systems. We expect this method to facilitate studies of high-redshift galaxies observed by interferometers. Additionally, we look forward to applying this image-plane lens modeling technique to strong lensing systems for dark matter research.

We have integrated the image-plane fitting method for interferometric data, as well as the pixelated source reconstruction module, into \texttt{Lenstronomy}. To facilitate reproducibility and usability, we provide example Jupyter notebooks demonstrating how to apply these methods, along with the relevant data files and scripts used to produce the results in this paper. All resources are available at: \url{https://github.com/nanz6/projects_of_strong_lensing/tree/main/lenstronomy_ALMA_likelihood_and_source_recon}.

\section*{Acknowledgements}
\label{sec ack}
This work has been supported by Brand Fortner and NSF AST-1715717. This work was partially supported by the Center for AstroPhysical Surveys (CAPS) at the National Center for Supercomputing Applications (NCSA), University of Illinois Urbana-Champaign.
SB is supported by the Department of Physics and Astronomy, Stony Brook University.

This paper makes use of the following ALMA data: ADS/JAO.ALMA \# 2013.1.00880.S, ADS/JAO.ALMA \# 2016.1.01293.S,  ADS/JAO.ALMA \# 2017.1.01423.S. ALMA is a partnership of ESO (representing its member states), NSF (USA) and NINS (Japan), together with NRC (Canada), MOST and ASIAA (Taiwan), and KASI (Republic of Korea), in cooperation with the Republic of Chile. The Joint ALMA Observatory is operated by ESO, AUI/NRAO and NAOJ.

\bibliography{main}
\bibliographystyle{aasjournal}
\appendix

\section{Appendix A: Definitions of Parametric Profiles}
\label{Sec param trans}
We briefly provide the definitions of the parametric lens and source profiles used in this paper, aiming to assist people working with \texttt{Lenstronomy} and \texttt{visilens} in understanding the parameter transformation between these two packages.

\subsection{\textup{A.1.} Lens Model: Singular Isothermal Ellipsoid}
The Singular Isothermal Ellipsoid (SIE) lens profile assumes that the galaxy consists of particles behaving like an isothermal ideal gas. For further reference, see \cite{narayan1996lectures} and \cite{kormann1994isothermal}. The convergence of SIE is
\begin{equation}
\kappa(x,y) = \frac{1}{2} \Big[\frac{\theta_E}{\sqrt{q(x-x_L)^2+(y-y_L)^2/q}}\Big],
\end{equation}
where $\theta_E$ is the Einstein radius. The parameter $q$ is the ratio of the minor and major axes of the ellipsoid and is thus bounded by $0<q\le1$. The coordinates $x$ and $y$ correspond to the directions along the major and minor axes, respectively.

For an ellipsoid, an alternative way to define ellipticity is in Cartesian coordinates $(e_1,e_2)$ in terms of the axis ratio $(q)$ and the position angle of the major axis $(\phi_L)$. In \texttt{Lenstronomy}, the ellipticity is defined by the following equations, as proposed by \cite{birrer2015gravitational}:
\begin{equation}
\label{ellipticity def in lenstronomy}
\begin{aligned}
e_1 &=\frac{1-q}{1+q}\cos(2\phi_L)\\
e_2 &=\frac{1-q}{1+q}\sin(2\phi_L)
\end{aligned}.
\end{equation}

For those familiar with \texttt{visilens}, the ellipticity of SIE in \texttt{visilens} is defined as
\begin{equation}
\label{ellipticity def}
e_{L,\rm{visilens}} = 1 - q.
\end{equation}
Thus, special care should be taken when comparing the ellipticities between \texttt{Lenstronomy} and \texttt{visilens}.

\subsection{\textup{A.2.} Lens External Shear}
\label{sec app external shear}
External shear summarizes the contribution to the lens shear field from structures that have not been modeled by the other lens density profiles. The external shear is specified by its strength, $\gamma$, and position angle, $\phi_\gamma$, or alternatively by $\gamma_1$ and $\gamma_2$ in Cartesian notation. These two parameterizations are related by
\begin{equation}
\begin{aligned}
\gamma_1 & = \gamma\cos(2\phi_\gamma),\\
\gamma_2 & = \gamma\sin(2\phi_\gamma).
\end{aligned}
\end{equation}
In addition to the strength and angle of the external shear, the position of the external shear also needs to be specified in computations. In \texttt{Lenstronomy}, the external shear center is specified by the two parameters $(\rm{ra}_0, \rm{dec}_0)$ for the ``external shear'' lens profile.

\subsubsection{\textup{A.2.1} Simultaneous Transformation of $(\rm{ra}_0, \rm{dec}_0)$ and the Center $(x_s,y_s)$ of Source Light}
We have introduced the center of the external shear, which is specified by the two parameters $(\rm{ra}_0, \rm{dec}_0)$ in \texttt{Lenstronomy}. However, the center of the external shear is not usually reported as free parameters in the strong lens modeling literature, because it is typically assumed that the center of the external shear coincides with the center of the main lens, $(x_L, y_L)$.

As \texttt{Lenstronomy} treats the center of the external shear $(\rm{ra}_0, \rm{dec}_0)$ independently from $(x_L, y_L)$, we need to determine how to transform $(\rm{ra}_0, \rm{dec}_0)$ from its value in \texttt{Lenstronomy} to $(x_L, y_L)$ whenever $(\rm{ra}_0, \rm{dec}_0) \neq (x_L, y_L)$, in order to compare fitting results with other packages.

This transformation is facilitated by the fact that the center of the external shear is completely degenerate with the source positions. Displacing $(\rm{ra}_0, \rm{dec}_0)$ leaves the lensed images unchanged if the source positions are simultaneously transformed as follows:
\begin{equation}
\label{trans external shear center}
\begin{aligned}
\rm{ra}_0 &\rightarrow \rm{ra}_{0}',\\
\rm{dec}_{0} &\rightarrow \rm{dec}_{0}',\\
x_s&\rightarrow x_s' = x_s + \gamma_1 (\rm{ra}_{0}' - \rm{ra}_0) + \gamma_2 (\rm{dec}_{0}' - \rm{dec}_0),\\
y_s&\rightarrow y_s' = y_s + \gamma_2 (\rm{ra}_{0}' - \rm{ra}_0) - \gamma_1 (\rm{dec}_{0}' - \rm{dec}_0).
\end{aligned}
\end{equation}
In this transformation, the center of the external shear is shifted from $(\rm{ra}_0, \rm{dec}_0)$ to a new position $(\rm{ra}'_0, \rm{dec}'_0)$, while the source positions are adjusted according to the strength of the external shear and the magnitude of the shift.

For example, suppose we want to compare the fitting results of \texttt{Lenstronomy} with those of \texttt{visilens}. In \texttt{visilens}, the center of the external shear is fixed at the center of the first SIE lens profile, i.e., $(\rm{ra}_0,\rm{dec}_0)_{visilens} =(x_L,y_L)$, while in \texttt{Lenstronomy}, $(\rm{ra}_0, \rm{dec}_0)$ is set to some constant during the fitting, usually $(\rm{ra}_0, \rm{dec}_0) = (0,0)$. After completing the \texttt{Lenstronomy} fitting, we transform the parameters as follows:
\begin{equation}
\label{trans external shear center1}
\begin{aligned}
\rm{ra}_0 &\rightarrow x_L,\\
\rm{ra}_{0} &\rightarrow y_L,\\
x_s&\rightarrow x_s' = x_s + \gamma_1 (x_L - \rm{ra}_0) + \gamma_2 (y_L - \rm{dec}_0),\\
y_s&\rightarrow y_s' = y_s + \gamma_2 (x_L - \rm{ra}_0) - \gamma_1 (y_L - \rm{dec}_0).
\end{aligned}
\end{equation}
After this transformation, the new source coordinates $(x_s',y_s')$ becomes consistent with their definition in \texttt{visilens}.

Furthermore, the source positions relative to the lens center under this transformation are given by
\begin{equation}
\label{source offset position}
\begin{aligned}
x_{\rm{source\ off}}=x_{s}' - x_L &= x_{s} - x_L + \gamma_1(x_L - \rm{ra}_0) + \gamma_2(y_L - \rm{dec}_0)\\
y_{\rm{source\ off}}=y_{s}' - y_L &= y_{s} - y_L + \gamma_2(x_L - \rm{ra}_0) - \gamma_1(y_L - \rm{dec}_0).\\
\end{aligned}
\end{equation}

\subsection{\textup{A.3.} Source Light Profile: Elliptical S\'{e}rsic}
An elliptical S\'{e}rsic light profile is defined as
\begin{equation}
I(R) = I_0\exp\Big\{-b_n\big[(R/R_{\rm{Sersic}})^{\frac{1}{n}}-1\big]\Big\},
\end{equation}
where $I_0$ is the $amp$ parameter defined in \texttt{Lenstronomy}, and $n$ is the S\'{e}rsic index, $n_{\rm{Sersic}}$. $b_n$ is approximately related to $n$ by $b_n = 1.999n - 0.327$, and $R$ is the effective distance from the center, $R=\sqrt{q(x-x_s)^2+(y-y_s)^2/q}$, with $q$ being the minor-to-major axis ratio. The source ellipticities $e_1$ and $e_2$ in \texttt{Lenstronomy} are related to the axis ratio in the same way as given by Equation (\ref{ellipticity def in lenstronomy}).

Note that in \cite{dong2019source}, the radial scale of the source profile is specified by the half-light radius of the major axis, which is related to the S\'{e}rsic radius by $R_{\rm{Major}}=R_{\rm{Sersic}}/\sqrt{q}$.

\section{Appendix B: Derivation of optimal regularization strength equation}
\label{Sec dri strength}
Just as described in \cite{mackay1992bayesian} and \cite{suyu2006bayesian}, the optimal regularization strength $\lambda_s$ can be obtained by exploiting the Bayes evidence $P(\boldsymbol{d}|\lambda_s)$ of Equation (\ref{bayes formula}). Recovering the normalization factor, we can write the prior function as
\begin{equation}
\label{appendix normalized bayes prior}
P(\{a_i\}|\lambda_s,U) = \frac{1}{Z_a(\lambda_s,U)}\exp(-\frac{1}{2}\lambda_s \sum_{ij}U_{ij}a_ia_j),
\end{equation}
where $Z_a(\lambda_s,U)$ is obtained by integrating over all parameters $\{a_i\}$
\begin{equation}
\label{appendix norm factor prior}
Z_{a}(\lambda_s,U) = \int e^{-\frac{1}{2}\lambda_s \sum_{ij}U_{ij}a_ia_j}d^{N_s}a=\frac{(2\pi)^{N_s/2}}{\sqrt{\det(\lambda_s U)}},
\end{equation}
where $N_s$ is the dimension of the source plane basis.
Similarly, the normalized likelihood is
\begin{equation}
\label{appendix normalized likelihood}
P(\boldsymbol{d}|\{a_i\}) = \frac{1}{Z_d}\exp\Big\{-\frac{1}{2} \big[\boldsymbol{d} - \mathbf{B}\boldsymbol{x}(a_i)\big]^T \mathbf{C}_{n}^{-1}\big[\boldsymbol{d} - \mathbf{B}\boldsymbol{x}(a_i)\big]\Big\},
\end{equation}
where the normalization factor $Z_d$ is obtained by integrating over $\boldsymbol{d}$. Hence it depends only on the covariance matrix $\mathbf{C}_n$. Applying the proportionality Equation (\ref{noise covariance psf proportionality}) and using the notation introduced by Equations (\ref{source recon b}) and (\ref{source recon m}), we can rewrite the product of the likelihood and the prior as
\begin{equation}
\label{appendix product likelihood and prior}
P(\boldsymbol{d}|\{a_i\})P(\{a_i\}|\lambda_s,U) = \frac{1}{Z_dZ_{a}(\lambda_s,U)}\exp\Big\{-\frac{1}{2}\big[
 \sum_{ij}(\lambda_s U_{ij} + M_{ij})a_ia_j
 - 2 \sum_i a_ib_i + \frac{\boldsymbol{d}^T\mathbf{B}^{-1}\boldsymbol{d}}{\sigma^2_n}
 \big] \Big\}.
\end{equation}
To get the Bayes evidence, we need to integrate over the linear coefficients $\{a_i\}$. Completing the square of the exponent and using the vector notation for $a$ and $b$, we can write the likelihood-prior product as
\begin{equation}
\label{appendix product likelihood and prior 1}
P(\boldsymbol{d}|\{a_i\})P(\{a_i\}|\lambda_s,U) = \frac{1}{Z_dZ_{a}(\lambda_s,U)}\exp\Big\{-\frac{1}{2}\big[
(a - \hat{a})^T(M+\lambda_s U)(a - \hat{a}) - b^T (M+\lambda_s U)^{-1}b
 + \frac{\boldsymbol{d}^T\mathbf{B}^{-1}\boldsymbol{d}}{\sigma^2_n}
 \big] \Big\},
\end{equation}
where $\hat{a} = (M+\lambda_s U)^{-1}b$, being the optimal linear coefficients under given regularization strength. Integrating over $\{a_i\}$ gives the marginalized posterior (also known as the Bayes evidence of the source reconstruction) as
\begin{equation}
\label{appendix bayes evidence}
P(\boldsymbol{d}|\lambda_s,U) = \frac{\exp(-\frac{\boldsymbol{d}^T\mathbf{B}^{-1}\boldsymbol{d}}{2\sigma^2_n})}{Z_d}\frac{\sqrt{\det(\lambda_s U)}}{\sqrt{\det(M+\lambda_s U)}}\exp\Big[\frac{1}{2}b^T(M+\lambda_s U)^{-1}b\Big],
\end{equation}
where we have used the specific form of the prior normalization Equation (\ref{appendix norm factor prior}).

Applying Bayes equation, the probability distribution of the regularization strength given data and regularization form can be written as
\begin{equation}
P(\lambda_s|\boldsymbol{d},U) = \frac{P(\lambda_s)}{P(\boldsymbol{d})}P(\boldsymbol{d}|\lambda_s,U).
\end{equation}
Using Equation (\ref{appendix bayes evidence}), the derivative of $\ln[P(\lambda_s|\boldsymbol{d},U)]$ with respect to $\lambda_s$ is
\begin{equation}
\label{appendix d p( lambda|data)}
\frac{\partial \ln P(\lambda_s|\boldsymbol{d},U)}{\partial \lambda_s} = \frac{d\ln P(\lambda_s)}{d\lambda_s} + \frac{1}{2}\Big\{ \frac{N_s}{\lambda_s}-{\rm{tr}}\big[(M+\lambda_sU)^{-1}U\big] - b^T (M+{\lambda}_sU)^{-1}U(M+{\lambda}_sU)^{-1}b \Big\}.
\end{equation}

The optimal regularization strength $\hat{\lambda}_s$ is the one maximizes $P(\lambda_s|\boldsymbol{d},U)$, which can be solved by $\frac{\partial \ln (P(\lambda_s|\boldsymbol{d}))}{\partial \lambda_s}|_{\lambda_s = \hat{\lambda}_s} = 0$, i.e.
\begin{equation}
\label{appendix d p( lambda|data) optimal}
0=\frac{d\ln P(\lambda_s)}{d\lambda_s}\Big|_{\lambda_s = \hat{\lambda}_s} + \frac{1}{2}\Big\{ \frac{N_s}{\hat{\lambda}_s}-{\rm{tr}}\big[(M+\hat{\lambda}_sU)^{-1}U\big] - b^T (M+\hat{\lambda}_sU)^{-1}U(M+\hat{\lambda}_sU)^{-1}b \Big\}.
\end{equation}
From this result, if we assume flat prior for the regularization strength, which means $\frac{d\ln P(\lambda_s)}{d\lambda_s}=0$, the equation solves $\hat{\lambda}_s$ is then
\begin{equation}
N_s = \hat{\lambda}_s {\rm{tr}}\big[(M+\hat{\lambda}_sU)^{-1}U\big] + \hat{\lambda}_s b^T (M+\hat{\lambda}_sU)^{-1}U(M+\hat{\lambda}_sU)^{-1}b.
\end{equation}
The above one is the equation that solves the optimal regularization strength in our natural weighting interferometric case, which corresponds to Eq.(20) of \cite{suyu2006bayesian}.

\section{Appendix C: Supplementary plots}
\label{sec appendix add plots}
\begin{figure*}[h!t]
\centering
  \epsscale{1.05}
  \plotone{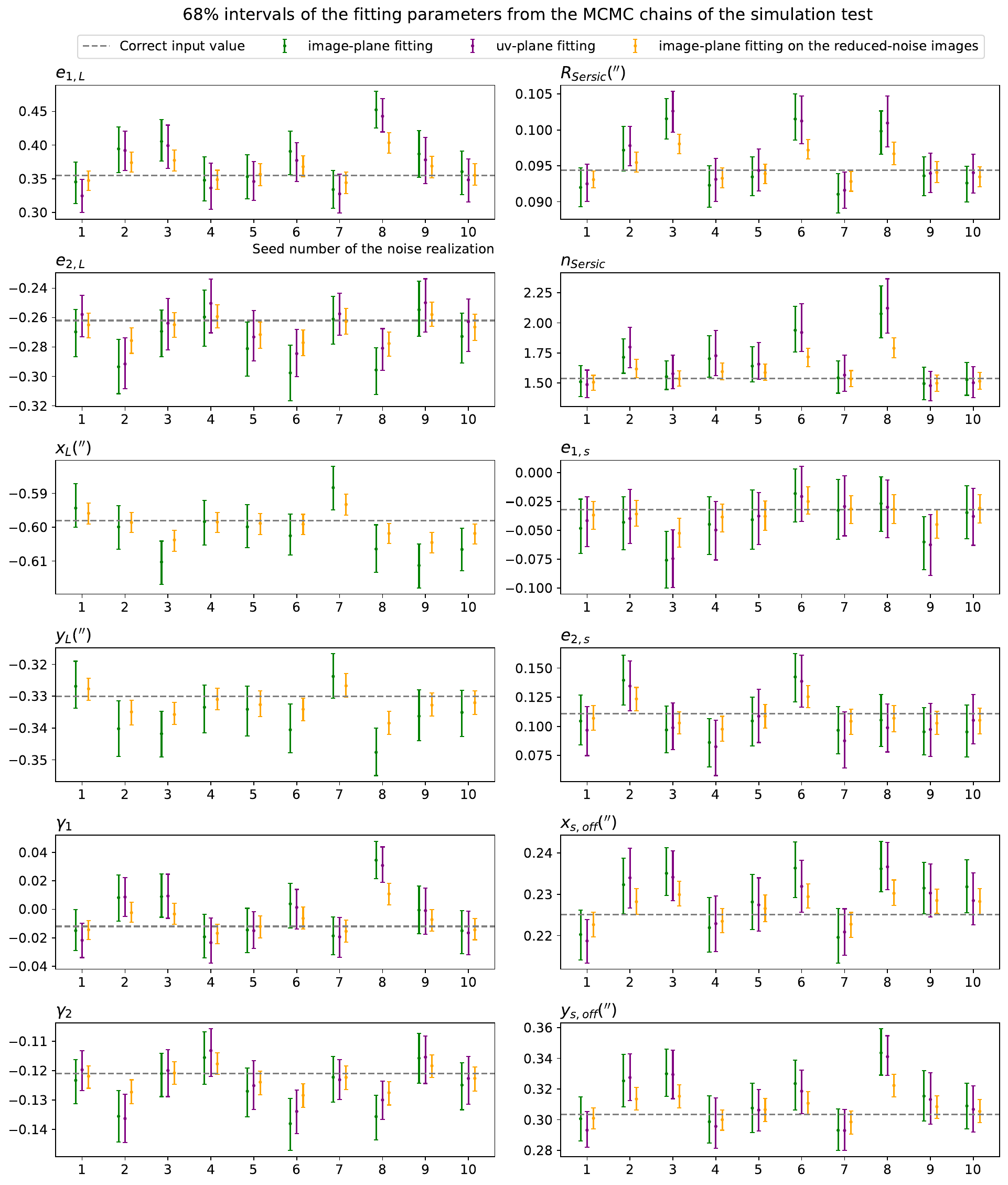}
  \caption{$1\sigma$ (68\%) intervals of parameters (except the Einstein radius) from the fittings in Test 1 and Test 2 in Section \ref{sec test likelihood}, continuing from Figure \ref{fig mcmc 1 sigma thetaE}. For the $uv$-plane fittings (purple vertical bars), the lens center parameters $x_L$ and $y_L$ are not plotted, as \texttt{visilens} uses a different definition of the center of the coordinate grids compared to \texttt{Lenstronomy}.}
  \label{fig mcmc 1 sigma others}
\end{figure*}

Figure \ref{fig mcmc 1 sigma others} is the continuation of Figure \ref{fig mcmc 1 sigma thetaE}.

\section{Appendix D: Regularization Matrix for Pixelated Source Reconstruction}
\label{sec appendix regularization}
In this section, we present our computational method for the zeroth-order, gradient, and curvature regularization matrices for square-gridded pixelated sources. The idea of the pixelated source regularization matrices is introduced in the Appendix of \cite{suyu2006bayesian}. We provide our computational method by writing the regularization matrices as block matrices.

Before constructing the regularization matrix, we need to label the pixels of the two-dimensional image with a single index to organize the source pixels into a one-dimensional vector. Note that in Section \ref{sec stat functions fitting} and Appendix \ref{Sec dri strength}, we have used $\{a_i\}$ to denote pixel values, where the pixel is labeled by the one-dimensional index $i$.

We assign the index $i$ to pixels by counting them starting from the smallest $x_{min}$ and $y_{min}$, then counting the pixels in an increasing $x$ sequence while keeping $y$ at $y_{min}$. After all pixels with $y=y_{min}$ have been counted, we continue with the pixels of $y = y_{min}+1$ and proceed in the sequence of increasing $x$. Assuming the rectangular source region has the size of $l_x \times l_y$ pixels, the Cartesian indices $(i_x, i_y)$ are related to the one-dimensional index $i$ by
\begin{equation}
i = (i_y - 1)l_x + i_x,
\end{equation}
where the Cartesian indices are integers within the ranges $1 \le i_x \le l_x$ and $1 \le i_y \le l_y$.

For the remainder of this section, we will use the indices $i$ and $i_x, i_y$ interchangeably.

\paragraph{Zeroth-order regularization}
The zeroth-order regularization matrix for a rectangular region of pixelated source is
\begin{equation}
\sum_{i,j} U_{ij}a_ia_j = \sum_{i}(a_i)^2.
\end{equation}
The zeroth-order matrix is simply the unity matrix
\begin{equation}
U= \mathbf{1}.
\end{equation}

\paragraph{Gradient regularization}
The gradient regularization term measures the first order difference between neighboring pixels. It is given by
\begin{equation}
\label{appendix gradient reg pixel}
\sum_{i,j} U_{ij}a_ia_j = \sum_{i_x=0}^{l_x}\sum_{i_y=0}^{l_y} \big[(a_{i_x,i_y} - a_{i_x+1,i_y})^2 +
(a_{i_x,i_y} - a_{i_x,i_y+1})^2\big].
\end{equation}

Note that we regard the pixels with 2d indices out of the domain in the above summation as the pixels outside of the source region for reconstruction. And we set their amplitudes to be zero. Symbolically, it means
\begin{equation}
a_{i_x,i_y}=0,\rm\ for\ i_x\notin\{1,...,l_x\}\rm\ or\  i_y\notin\{1,...,l_y\}.
\end{equation}
In this way, we are assuming the region out of the source reconstruction region with zero flux and the reconstruction is therefore on a blank background.

From the definition, we observe that the gradient regularization matrix $U$ can be written in a repeated pattern as a block matrix
\begin{equation}
U=\underbrace{\left(\begin{array}{ccccccc}
\mathbf{u}_1 & \mathbf{-1} &  &  & & &\\
\mathbf{-1} & \mathbf{u}_1 & \mathbf{-1} &  & & & \\
&\mathbf{-1} & \mathbf{u}_1 &  &  & & \\
 & & & \ddots & &  & \\
&&&& \mathbf{u}_1 & \mathbf{-1} &  \\
&&&& \mathbf{-1} & \mathbf{u}_1 & \mathbf{-1}\\
&&&& &\mathbf{-1} & \mathbf{u}_1
\end{array}\right)}_{l_y\times l_y\rm\ blocks},{\rm\ with} \
\mathbf{u}_1=\underbrace{\left(\begin{array}{ccccccc}
4 & -1 &  &  & & &\\
-1 & 4 & -1 &  & & & \\
&-1 & 4 &  &  & & \\
 & & & \ddots & &  & \\
&&&& 4 & -1 &  \\
&&&& -1 & 4 & -1\\
&&&& &-1 & 4
\end{array}\right)}_{l_x\times l_x}.
\end{equation}
All omitted entries are zero and the $\mathbf{-1}$ on the left is $-1$ times a unity matrix with dimension $l_x\times l_x$. The above equation gives a way to effectively compute the gradient regularization matrix.

\paragraph{Curvature regularization}
Curvature regularization measures the second order derivatives among pixels. It is given as
\begin{equation}
\sum_{i,j} U_{ij}a_ia_j = \sum_{i_x=0}^{l_x+1}\sum_{i_y=0}^{l_y+1} \big[(2a_{i_x,i_y} - a_{i_x+1,i_y} - a_{i_x-1,i_y})^2 +
(2a_{i_x,i_y} - a_{i_x,i_y+1} - a_{i_x,i_y-1})^2\big].
\end{equation}
In the same way discussed in the gradient regularization case, for the indices running outside of the source region, we treat the pixels to be zero. The curvature regularization matrix also has a repeated block matrix pattern
\begin{equation}
U=\underbrace{\left(\begin{array}{ccccccccc}
\mathbf{u}_2 & \mathbf{-4} & \mathbf{1} &  & & & & &\\
\mathbf{-4} & \mathbf{u}_2 & \mathbf{-4} & \mathbf{1} & & & & &\\
\mathbf{1} & \mathbf{-4} & \mathbf{u}_2 & \mathbf{-4} &  & & & &  \\
&\mathbf{1} & \mathbf{-4} & \mathbf{u}_2  &  & & & &  \\
 & & & &\ddots & & & & \\
 &&&&&\mathbf{u}_2 & \mathbf{-4} & \mathbf{1} &\\
 &&&&&\mathbf{-4} & \mathbf{u}_2 & \mathbf{-4} & \mathbf{1}\\
 &&&&&\mathbf{1} & \mathbf{-4} & \mathbf{u}_2 & \mathbf{-4}\\
 &&&&&&\mathbf{1} & \mathbf{-4} & \mathbf{u}_2\\
\end{array}\right)}_{l_y\times l_y\rm\ blocks},{\rm\ with} \
\mathbf{u}_2=\underbrace{\left(\begin{array}{ccccccccc}
12 & -4 & 1 &  & & & &&\\
-4 & 12 & -4 & 1 & & & && \\
1&-4 & 12 & -4 & && & & \\
&1&-4 & 12 &  & & & & \\
 & & & & \ddots & & & & \\
&&&&&  12 & -4 & 1 &\\
&&&&& -4& 12 & -4 & 1 \\
&&&&&1& -4 & 12 & -4\\
&&&&& & 1 &-4 & 12
\end{array}\right)}_{l_x\times l_x}.
\end{equation}
All omitted entries are zero.

\end{document}